\documentclass{nature}
\usepackage{amsmath,amssymb,enumerate}
\usepackage{lineno}
\usepackage{graphicx,float,color}
\usepackage[font={small,singlespacing},labelfont=bf]{caption}
\DeclareCaptionLabelSeparator{vline}{~$\big\rvert$ }
\DeclareCaptionLabelFormat{extended}{Extended Data Fig. #2}
\DeclareCaptionLabelFormat{supplementary}{Supplementary Fig. #2}
\usepackage{txfonts}
\usepackage{comment}
\usepackage[dvipsnames]{xcolor}
\usepackage[colorlinks,citecolor=blue,linkcolor=magenta]{hyperref}
\usepackage{textcomp}
\usepackage{siunitx}

\newlength{\boxwidth}
\def\btheorem{\begin{theorem}}
\def\etheorem{\end{theorem}}
\def\blemma{\begin{lemma}}
\def\elemma{\end{lemma}}
\def\bproposition{\begin{proposition}}
\def\eproposition{\end{proposition}}
\def\bcorollary{\begin{corollary}}
\def\ecorollary{\end{corollary}}
\def\bdefinition{\begin{definition}}
\def\edefinition{\end{definition}}
\def\bexample{\begin{example}}
\def\eexample{\end{example}}
\def\bremark{\begin{remark}}
\def\eremark{\end{remark}}

\newcommand{\be}{\begin{equation}}
\newcommand{\ee}{\end{equation}}
\newcommand{\beq}{\begin{eqnarray}}
\newcommand{\eeq}{\end{eqnarray}}
\newcommand{\bem}{\begin{multline}}
\newcommand{\eem}{\end{multline}}
\newcommand{\ba}{\begin{align}}
\newcommand{\ea}{\end{align}}

\title{Hierarchical Granular Metamaterials}

\author{James Utama Surjadi$^{1,\dagger}$, Bastien F. G. Aymon$^{1,\dagger}$, Ayan Kumar$^{1}$, Lei Wu$^{1}$, Jet Lem$^{1}$, Ken N. Kamrin$^{1,2}$, Carlos M.~Portela$^{1\ast}$}
\begin{document}
\maketitle

\begin{affiliations}
\footnotesize
 \item Department of Mechanical Engineering, Massachusetts Institute of Technology, Cambridge, MA 02139, USA
 \item Department of Mechanical Engineering, University of California Berkeley, Berkeley, CA 94720, USA
  \item[$^\dagger$] Equal contribution 
  \item[$^\ast$] e-mail: cportela@mit.edu 
\end{affiliations}
\vspace{10pt}
% \linenumbers
\spacing{1.0}

\begin{abstract}
Granular materials dissipate energy efficiently through intergranular interactions, yet their disordered, dense nature precludes precise control and integration into lightweight systems. Architected materials offer tunable mechanical responses at low densities but tend to localize stress, limiting dissipation efficiency. Here, we introduce hierarchical granular metamaterials that reconcile these trade-offs through three levels of design: lightweight architected grains engineered with hollow elliptical inclusions,  crystal-inspired grain packings, and functional gradients and defects within grain tessellations. These metamaterials exhibit simultaneous increases in impact energy absorption per unit mass and reductions in transmitted peak force at low densities, outperforming conventional architected materials. \textit{In situ} nanomechanical experiments and nonlinear computational models reveal that enhanced lateral grain expansion drives recruitment of neighboring grains, amplifying plastic and frictional dissipation. Multiscale impact experiments confirm that these mechanisms persist across length scales, constituent materials, and dimensionalities. Beyond mechanical performance, we demonstrate that spatially programmable inter-grain contact networks enable deterministic routing of deformation, which extends to electrical transport pathways independently of packing geometry. By combining granular principles with architected material design, this work establishes a paradigm for multifunctional metamaterials whose contact topology, mechanical response, and transport properties can be programmed independently.

\end{abstract}
\newpage

\subsection{Main}
\hfill\\
Granular materials are ubiquitous and are widely used for impact dissipators (e.g., sandbag blast mitigation) due to their ability to quickly redistribute local impact forces through branching grain-to-grain contact interactions\cite{jaeger1996granular, majmudar2005contact, Clark_Petersen_Kondic_Behringer_2015}. Accordingly, this mechanism underlies applications ranging from impact-absorbing barriers and blast mitigation to helmets and pads, packaging media, and vibration isolation. However, the dense, disordered nature of granular packings makes their mechanical response challenging to tune and predict. Consequently, most predictive frameworks to date employ homogenization assumptions\cite{Dunatunga_Kamrin_2017} or empirically calibrated scaling laws\cite{Clark_Behringer_2013, Katsuragi_Durian_2007}, which limits quantitative design control and transferability across loading regimes. Furthermore, the heterogeneous force-chain networks in random packings concentrate load transmission along a subset of contacts and grains, leaving considerable portions of the packing only weakly engaged in load bearing\cite{Radjai_Wolf_Jean_Moreau_1998}.

When lightweight performance is required, foams provide an alternative route to impact dissipation\cite{deshpandeFleck2000foams, barnes2014dynamic}, but their stochastic cellular architecture similarly limits precise mechanical control. Over the past decade, architected materials have emerged as a deterministic successor to foams, offering ultralight structures\cite{schaedler2011ultralight} with high stiffness and strength \cite{bauer2016approaching, zhang2019lightweight, crook2020plate}, resilience\cite{zheng2014ultralight, meza2014strong, portela2020extreme}, extreme stretchability\cite{surjadi2025double, carton2026design, cline2026entanglement, yan2020soft} and high toughness \cite{guell2019ultrahigh, surjadi2025double} that considerably outperforms their stochastic foam counterparts at similar relative densities (i.e., volume fraction)\cite{ bauer2017nanolattices, surjadi2025enabling}. Under dynamic loading, tailored lattices exhibit enhanced energy absorption and tailored stress-strain responses through buckling-mediated mechanisms\cite{schaedler2014designing, frenzel2016tailored, clough2019elastomeric} or compaction fronts \cite{weeks2023effect}, and impact-mitigating realizations approach the per-unit-mass performance of conventional protective materials\cite{Portela_2021, butruille2024decoupling, surjadi2025exploiting}. Nevertheless, under localized high-rate loading such as ballistic impact, architected materials tend to localize deformation near the impact site, leading to collapse and densification that confines energy dissipation to a limited region\cite{Portela_2021, butruille2024decoupling, surjadi2025exploiting}. This inherent localization remains a key limitation for lightweight, energy-dissipating cellular materials.

Efforts to bridge these two paradigms have given rise to granular metamaterials---assemblies of discrete grains or building blocks that integrate the architectural tunability of metamaterials with the collective, nonlinear dynamics of granular media. This direction builds upon ordered granular crystals, which established that contact nonlinearity and packing topology can guide solitary waves, directional wave propagation, and impact-mitigation pathways\cite{Daraio_Nesterenko_Herbold_Jin_2005, Leonard_Fraternali_Daraio_2011, Burgoyne_Newman_Jackson_Daraio_2015, Burgoyne_Daraio_2014, Burgoyne_Daraio_2015}.
Building on these foundations, recent advances have expanded granular metamaterials beyond simple spherical particles through the design of particle geometry and inter-particle interactions\cite{athanassiadisJaeger2014}. Shape-engineered and tessellated grains have been shown to produce emergent mechanical responses, including auxeticity\cite{gaspar2010granular, haver2024elasticity} and tunable elasticity\cite{pashine2023tessellated} and plasticity \cite{karuriya2023granular}. Complementary approaches have focused on introducing reconfigurability into granular assemblies, using entangled or interlocked building blocks\cite{wang2021structured, zhou20253d, pezeshki2025tunable}, dynamic bonding schemes\cite{meng2024granular}, and beaded architectures\cite{dreier2025beaded} to achieve programmable stiffness, reversible shape transformation and adaptive mechanical behavior. More broadly, these concepts have enabled granular systems designed for reusable energy absorption\cite{fu2019programmable}, load-bearing architected materials\cite{dierichs2021designing, karuriya2023granular}, and collective robotic functionalities\cite{yang2025electronic}. Nevertheless, most studies to date have been limited to macroscale assemblies, where tunability is achieved primarily by varying grain geometry, packing arrangement, or confinement conditions. This limitation becomes particularly critical in ballistic and impact-mitigation applications, where the impactor size can approach or even fall below that of individual grains---failing to achieve separation of scales and thereby reducing collective interactions and energy dissipation efficiency. Moreover, existing granular metamaterials are typically arranged in random or periodic packings, which further constrains their design space, scalability, and applicability under extreme dynamic loading.

Here, we report a hierarchical design strategy that proceeds from architected grains to ordered packings, spatial heterogeneity, and programmable contact toward tunable properties across length and time scales. We first harness architected grains with precisely tunable density and anisotropy to control single-grain compliance, lateral expansion, and force-chain directionality. We then program their collective response by assembling them into crystalline packings and introducing crystal-inspired motifs---including interstitials, vacancies, and functional gradients---to deliberately perturb force-transmission pathways. This multiscale architecture enables synergistic interactions between intragranular deformation and intergranular rearrangement, enhancing energy dissipation while suppressing transmitted peak forces at ultralow density. Finally, we expand this framework by incorporating spatially varying connecting struts between architected grains, thereby prescribing programmable contact pathways. These inter-grain connections regulate not only stress propagation but also interfacial contact area and connectivity, enabling localized modulation of electrical transport properties. By bridging the discrete mechanics of granular media with architected materials design, this approach establishes a unified platform in which geometry governs both force transmission and functional transport. This design framework opens avenues for structurally robust thermal management systems, impact-adaptive sensing platforms, and energy-dissipative architectures for electronics, aerospace, and protective technologies.

\begin{figure}[h!]
    \centering
    \includegraphics[width=0.9\textwidth]{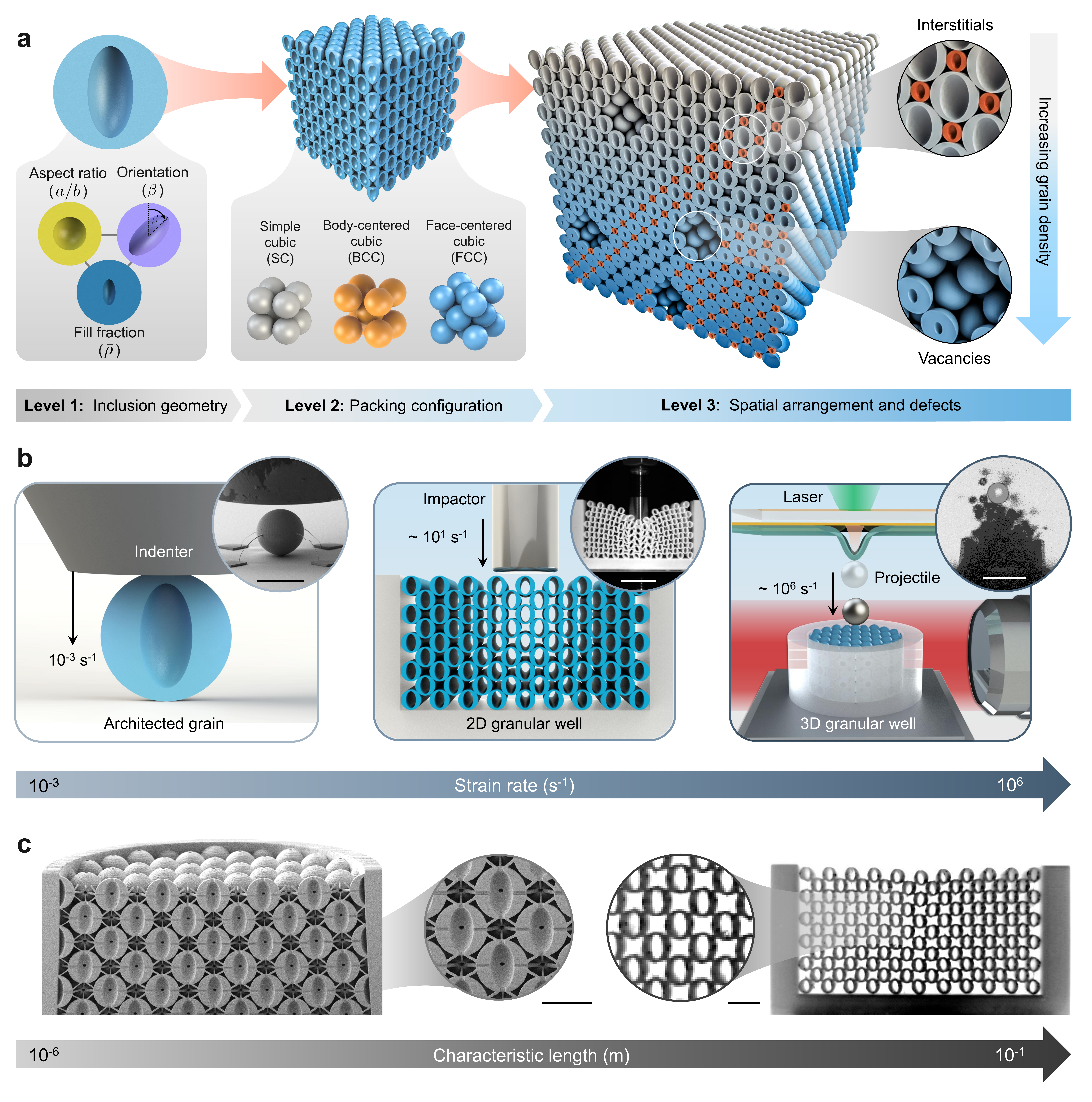}
    \caption{\textbf{Hierarchical granular metamaterials across length and time scales.} \textbf{a}, Schematic of the design paradigm for hierarchical granular metamaterials. \textbf{b}, Mechanical experiments conducted across a range of strain rates. Inset: representative snapshot of fabricated samples under load. Scale bars from left to right, 60 \textmu{}m, 25 mm, and 100 \textmu{}m. \textbf{c}, Fabricated hierarchical granular metamaterials at the micro- and macroscale. Microscale metamaterials show supporting sacrificial struts that are removed prior to characterization. Scale bars, 20 \textmu{}m (microscale), 10 mm (macroscale).}
    \label{fig:fig1_design}
\end{figure}

\subsection{Design of granular metamaterials with hierarchical architectures}
\hfill\\
To achieve programmable impact dissipation across scales, we developed a granular metamaterial design framework comprising three hierarchical architecture levels (Fig.~\ref{fig:fig1_design}\textbf{a}). At the grain level, each unit contains a hollow elliptical inclusion defined by its aspect ratio ($a/b$), orientation ($\beta$), and relative density ($\bar{\rho}$). These parameters collectively control the grain’s anisotropic stiffness and deformation under load (Extended Data Fig. 1 and Supplementary Note 1). At the assembly level, the architected grains are organized into periodic packings such as simple cubic (SC) and face-centered cubic (FCC) lattices. These configurations dictate interparticle contact geometry and influence the onset of collective rearrangement during deformation. Lastly, at the third level and beyond periodic packings, we introduce crystal-inspired spatial motifs---including functional gradients, vacancies, and interstitials---to embed controlled heterogeneity and enhance energy dissipation.

To investigate the mechanics governing these granular metamaterials at each design level, we performed quasi-static and dynamic impact experiments on both 2D (hollow cylindrical) and 3D (spherical) packings. The experiments spanned multiple length scales and included \emph{in situ} nanomechanical compressions, drop tower experiments, and laser-induced particle impact testing (LIPIT) (Fig.~\ref{fig:fig1_design}\textbf{b} and \textbf{c}). A confining wall contained the grains during loading and ensured consistent boundary conditions, while sacrificial struts between adjacent grains preserved the initial packing geometry prior to loading (Extended Data Fig. 2 and Supplementary Note 2). Together, these experiments directly reveal grain deformation, rotation, and contact evolution across scales, establishing the mechanistic basis for enhanced impact dissipation, with simulations used to interpret and quantify the underlying dissipation mechanisms.

\subsection{Tunable anisotropy and force chain manipulation via architected grains}
\begin{figure}[h!]
    \centering
    \includegraphics[width=0.87\textwidth]{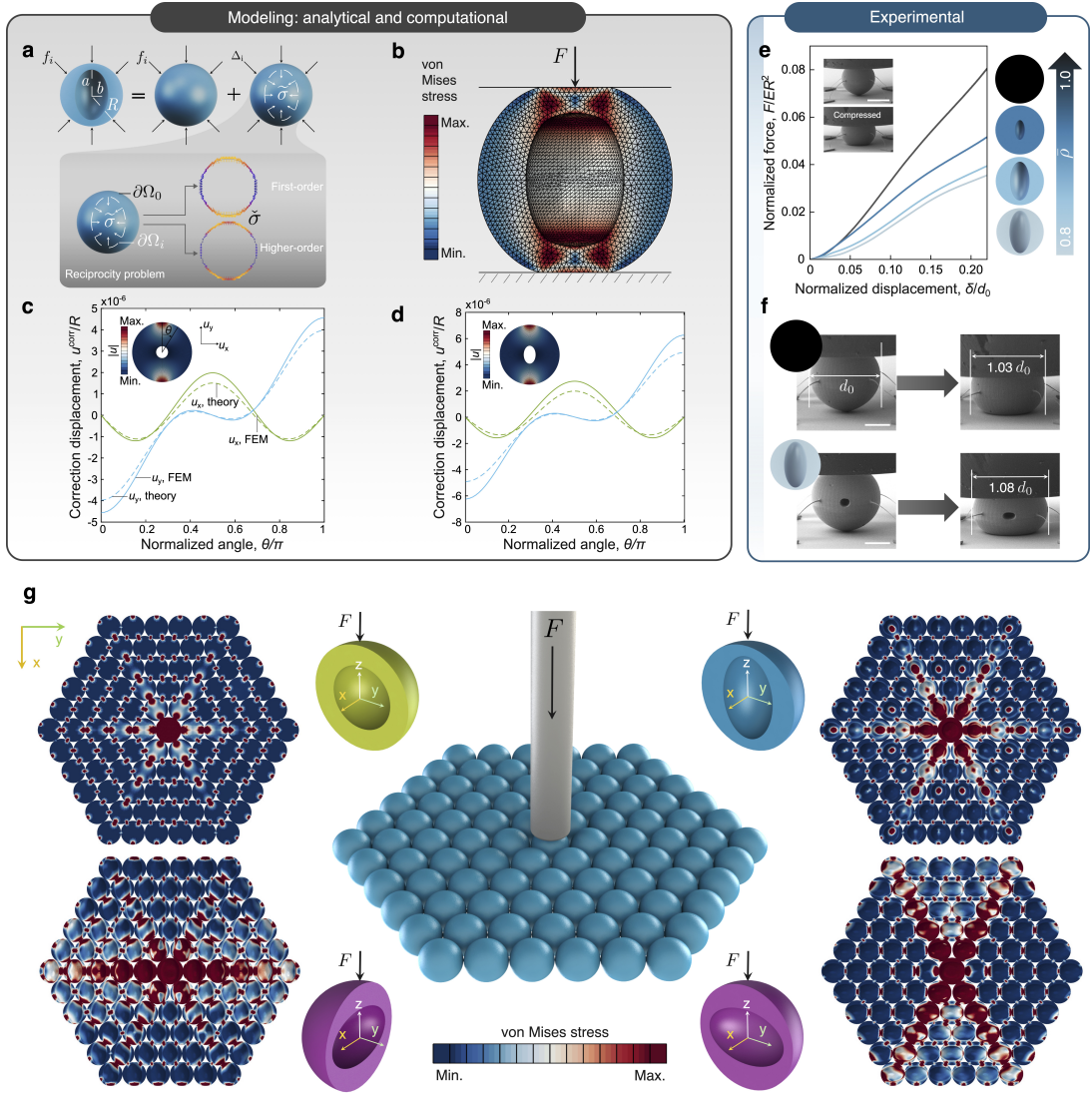}
    \caption{\textbf{Tunable mechanical response of architected grains.} 
\textbf{a}, Analytical decomposition of the compression problem for a spherical grain containing an ellipsoidal void. The architected-grain response is written as the response of a fully dense sphere plus a correction field associated with the void boundary, which is evaluated using a reciprocity problem with first- and higher-order terms. \textbf{b}, Stress distribution in an architected grain ($a/b$ = 1.5, $\beta = 0^{\circ}$, $\bar{\rho}$ = 81\%) upon compression obtained via FEM at 10\% strain. \textbf{c, d}, Normalized correction displacement, $u^{\mathrm{corr}}/R$, as a function of normalized angle, $\theta/\pi$, comparing analytical predictions for $u_x$ and $u_y$ with FEM; inset shows the displacement magnitude, $|u|$. \textbf{e}, Experimental normalized force--displacement curves, $F/ER^2$ versus $\delta/d_{0}$, for fully dense and architected grains with relative density, $\bar{\rho}$, decreasing from 1.0 to 0.8; inset shows representative compression images. \textbf{f}, Experimental comparison of lateral expansion in a fully dense grain and an architected grain under matched compression. \textbf{g}, Simulated von Mises stress fields in a single-layer architected granular packing loaded by a central indenter, showing tunable force-chain propagation for different void orientations.}
    \label{fig:fig2_grains}
\end{figure}

\hfill\\
To elucidate the mechanics underlying the deformation of the architected grains, we developed a theoretical framework based on Eshelby’s inclusion theory (Fig.~\ref{fig:fig2_grains}\textbf{a}) to describe the linear-elastic force–displacement response of grains containing elliptical voids (Supplementary Note 3). By solving the resulting reciprocity problem for rotationally symmetric configurations, we isolated the influence of inclusion shape and orientation on the overall deformation field. We additionally performed nonlinear FEM simulations incorporating frictional contact at the loading interfaces to capture the response of the complete grain under compression and the resulting stress distribution (Fig.~\ref{fig:fig2_grains}\textbf{b}). The constituent material properties were determined from independent compression experiments of the printed polymer (Supplementary Fig. 1). The predicted correction displacements along the outer grain surface, $u^{\text{corr}}$---defined as the void-induced difference between the displacement fields of the architected grain and an otherwise identical fully dense grain---agreed with the FEM results for representative void geometries, validating the analytical treatment of the deformation field (Fig.~\ref{fig:fig2_grains}\textbf{c,d}). The combined analytical and computational results predict an approximately linear decrease in effective stiffness with increasing porosity (Supplementary Note 3). They reveal that lower-density grains, achieved via the elliptical voids, exhibit both larger axial and lateral deformation under compression, demonstrating direct control of the grain mechanics through the architected inclusion. 

To further investigate the mechanical response of architected grains containing elliptical inclusions and confirm the theory's findings, we employed two-photon polymerization (TPP) to enable sub-micron fidelity in both grain geometry and internal architecture. The fabricated grains, composed of IP-Dip photoresist (\emph{Methods}), were characterized by \emph{in situ} compression inside a scanning electron microscope (SEM) to directly visualize their deformation. The stiffness of the grains increased with relative density, as expected from the higher solid fraction (Fig.~\ref{fig:fig2_grains}\textbf{e}). A similar trend was observed when varying the inclusion geometry (Extended Data Fig. 3 and Supplementary Note 4). Notably, when compressed to the same normalized displacement, lower-density grains exhibited substantially greater lateral deformation (Fig.~\ref{fig:fig2_grains}\textbf{f} and Supplementary Video 1), revealing a strong coupling between axial compliance and lateral expansion and further underscoring the role of internal architecture in governing grain-level mechanics. This controlled grain-level behavior is promising, as grain compliance has previously been shown to control nonlinear force propagation and enhance mass recruitment in traditional granular packings\cite{Clark_Petersen_Kondic_Behringer_2015}. While the analytical and computational models explore the linear-elastic response of these architected grains to small strains, the  experiments confirmed the expected trend that grains with lower axial stiffness undergo greater lateral expansion under compression, even at moderate strains where the constituent material undergoes inelastic deformation.

To demonstrate the role of internal grain architecture on the resulting intergranular interactions, we modeled a 2D hexagonal close-packed arrangement of grains confined by rigid walls and indented at the center of the packing from above (Fig.~\ref{fig:fig2_grains}\textbf{g}). Through variation of inclusion parameters (and orientation), the model demonstrates that anisotropic grain architecture can be harnessed by varying the internal inclusion geometry to tune the magnitude and directionality of force chains propagating through the network. Together, these results establish architected grains as programmable mechanical building blocks where microscale geometry governs both individual compliance and collective force transmission---laying the groundwork for granular metamaterials with tunable mechanical behavior. 

\subsection{Microscale hierarchical granular metamaterials across strain rates
}
\begin{figure}[h!]
    \centering
    \includegraphics[width=1.0\textwidth]{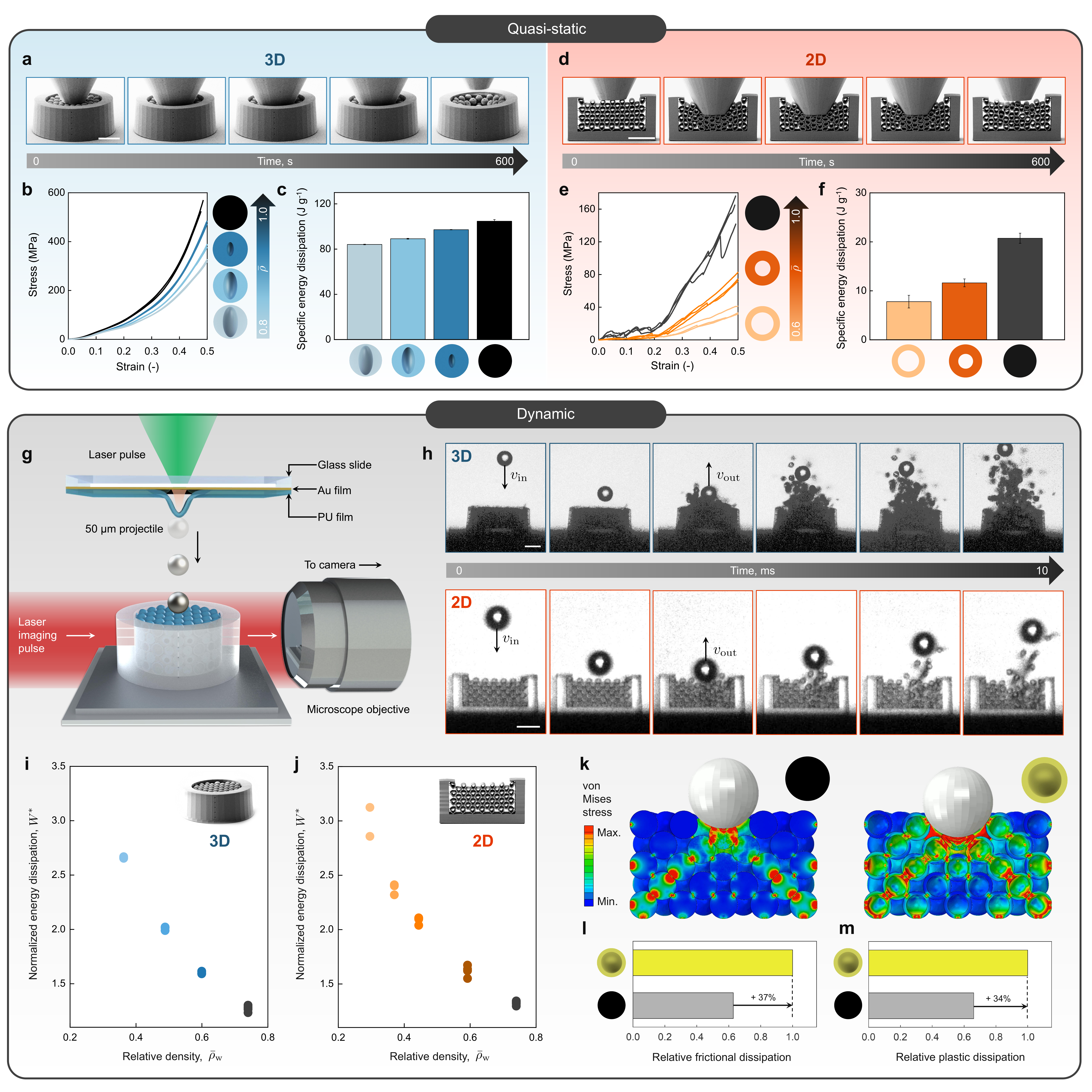}
    \caption{\textbf{Quasi-static and dynamic mechanical response of architected granular packings.} \textbf{a} and \textbf{d}, \emph{In situ} SEM snapshots of 3D and 2D architected granular packings under quasi-static indentation. \textbf{b} and \textbf{e}, Corresponding engineering stress–strain curves of 3D and 2D granular packings at various relative densities. \textbf{c} and \textbf{f}, Specific energy dissipation of 3D and 2D granular packings at various relative densities. \textbf{g}, Schematic of the LIPIT setup used to probe high-rate impact response. \textbf{h}, High-speed camera snapshots of silica microparticle impact on 3D and 2D architected granular packings. \textbf{i} and \textbf{j}, Normalized impact energy dissipation measured across 3D and 2D packings of varying grain densities. \textbf{k}, Nonlinear finite element simulations of impact demonstrating enhanced grain recruitment in low-density, architected packings. \textbf{l} and \textbf{m}, Quantitative comparison of relative frictional and plastic dissipation per unit mass in packings with architected and fully dense grains.}
    \label{fig:fig3_packing}
\end{figure}

\hfill\\
To determine the role of grain-level architecture on the mechanical behavior of the packing, we assembled architected grains into an ordered close packing configuration with containment walls (in 2D and 3D). During \emph{in situ} quasi-static indentation, both 2D and 3D packings exhibited pronounced grain deformation and rearrangement, with lateral motion and rotation particularly evident in the 2D configuration (Fig.~\ref{fig:fig3_packing}\textbf{a} and \textbf{d}). The corresponding engineering stress–strain curves showed progressive stiffening with increasing grain relative density, consistent with single-grain results (Fig.~\ref{fig:fig3_packing}\textbf{b} and \textbf{e}). Varying other inclusion geometry parameters (Supplementary Note 5), such as aspect ratio ($a/b$) and orientation ($\beta$), had a comparatively minor effect on the stress–strain response of the packings (Extended Data Fig. 4). To assess energy dissipation, we quantified the specific energy absorption by integrating the stress-strain response normalized by the absolute density of each sample (Fig.~\ref{fig:fig3_packing}\textbf{c} and \textbf{f}). As the inclusion increased in volume, thus lowering the density of the packing, specific energy dissipation correspondingly decreased. This response mirrors trends observed in conventional architected materials and stochastic foams, indicating that porosity within grains does not inherently enhance energy dissipation under quasi-static loading.

Given that granular media are most effective under dynamic loading, we examined the high-rate response of our packings using LIPIT (Fig.~\ref{fig:fig3_packing}\textbf{g}). In these experiments, we selected the projectiles to be 50 \textmu{m}-diameter silica microparticles, accelerated toward the sample at speeds ranging from \qty{180}{\metre\per\second} to \qty{270}{\metre\per\second} for 3D packings, or \qty{30}{\metre\per\second} to \qty{60}{\metre\per\second} for 2D packings, while a high-speed camera captured frames from the impact and rebound (\emph{Methods}). From the particle trajectories (Fig.~\ref{fig:fig3_packing}\textbf{h}), we extracted incident and rebound velocities to determine the change in kinetic energy before and after impact (Extended Data Fig. 5). This metric was then used to calculate the normalized energy dissipation ($W^*=\Delta T/\bar{\rho}T_0$, where $\Delta T$ is the kinetic energy lost by the projectile and $T_0$ the initial projectile kinetic energy) across packings of varying relative densities (i.e., fill fractions). In contrast to the quasi-static results, lower-density architected granular packings exhibited greater normalized energy dissipation (Fig.~\ref{fig:fig3_packing}\textbf{i} and \textbf{j}), demonstrating that internal grain architecture enables both reduced density and more efficient impact attenuation at high strain rates. This behavior is consistent with prior observations in stochastic granular impact, where the depth and efficiency of penetration depend sensitively on the initial packing state and the ability of the medium to rearrange contacts during loading\cite{umbanhowar2010granular}. In our architected system, however, this rearrangement is no longer stochastic but is geometrically encoded by the grain architecture and packing configuration itself.

To identify the mechanisms behind this response, we performed nonlinear finite element simulations incorporating frictional contact and material nonlinearity (\emph{Methods}). The simulations revealed that more compliant architected grains enhanced grain recruitment during impact, promoting collective deformation and strain energy redistribution throughout the packing (Fig.~\ref{fig:fig3_packing}\textbf{k}). Quantitative analysis further showed that assemblies of architected grains exhibited greater frictional and plastic dissipation per unit mass than those composed of fully dense grains (Fig.~\ref{fig:fig3_packing}\textbf{l} and \textbf{m}). These trends are consistent with collisional models of granular impact, in which energy dissipation scales with the number of activated grain–grain contacts and the magnitude of relative motion at each contact\cite{bester2017collisional}. By increasing lateral grain expansion, our architected grains effectively raise both quantities simultaneously. These results demonstrate that microscale grain architecture fundamentally governs how granular packings distribute and dissipate energy, establishing a pathway toward lightweight granular metamaterials that achieve enhanced impact mitigation through hierarchical design.

\subsection{Programmable dissipation via defective packings}
\begin{figure}[H]
    \centering
    \includegraphics[width=1\textwidth]{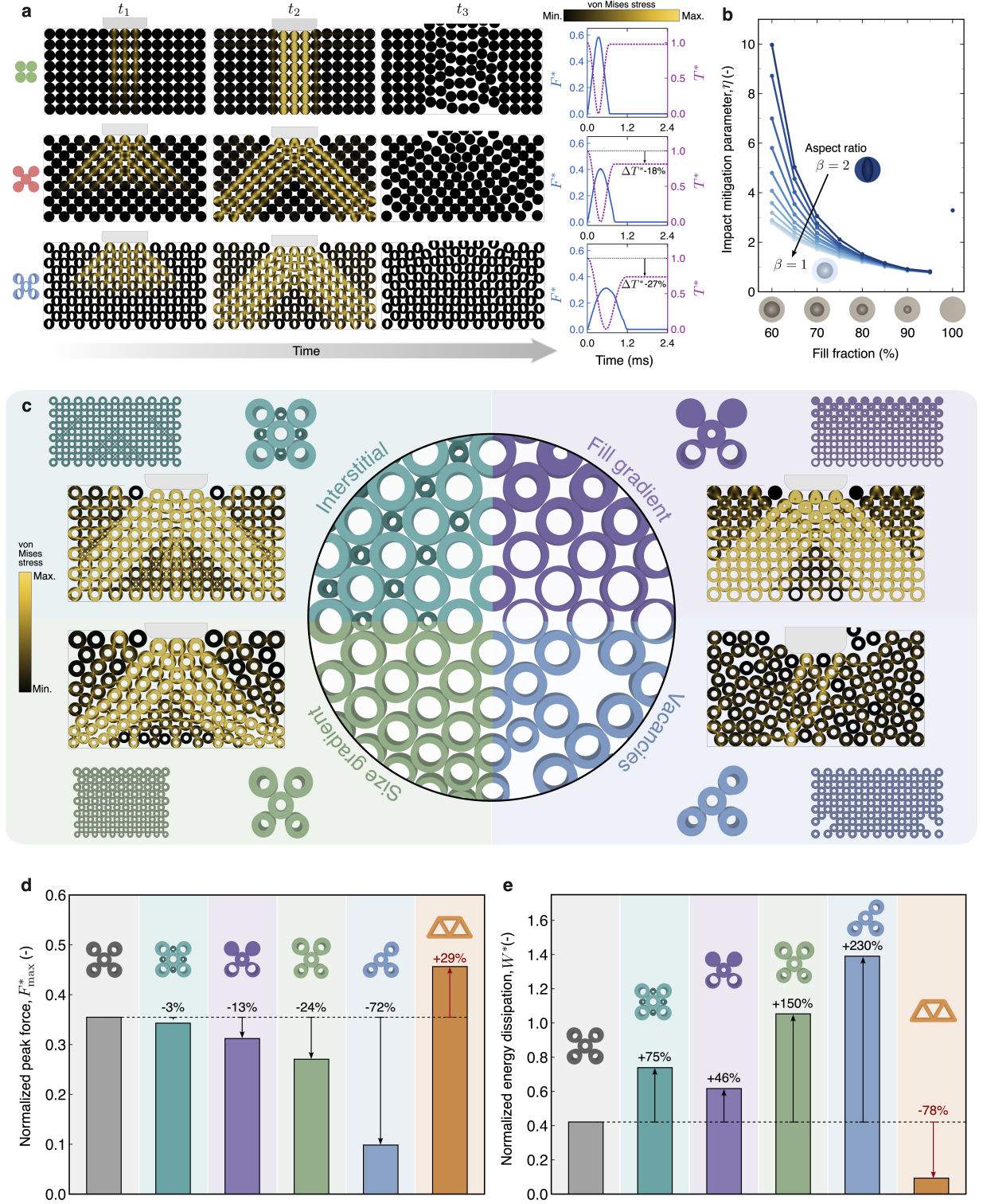}
    \caption{\textbf{Hierarchical design principles for programmable granular metamaterials.} \textbf{a}, Finite element model (FEM) results comparing force propagation and spatiotemporal response in simple cubic (SC) and close packings with and without architected grains. \textbf{b}, Parametric study on the effect of inclusion geometry (aspect ratio and relative density) on impact mitigation parameter. \textbf{c}, FEM results presenting the effect of defects on the packing behavior. \textbf{d} and \textbf{e}, Normalized peak force and energy dissipation of different defect configurations showing significant improvement compared to a control lattice structure.}
    \label{fig:fig4_simulations}
\end{figure}

\hfill\\
To further understand and program energy dissipation mechanisms in hierarchical granular metamaterials, we developed and experimentally validated an FEM framework capable of predicting the response of 2D architected granular packings under indentation (Fig.~\ref{fig:fig4_simulations} and Extended Data Fig. 6). This framework enabled systematic exploration of design parameters, spanning packing configurations, inclusion geometries, and relative densities.

We first examined the influence of packing configuration on the resulting force patterns (Fig.~\ref{fig:fig4_simulations}\textbf{a}). Simulations revealed that varying the spatial arrangement from SC to close packings greatly alters force redistribution pathways. In the SC packing, force transmission remains confined to vertical chains aligned with the loading direction, limiting stress delocalization and restricting load sharing to a small number of grains. In contrast, the close packing generated angled force chains that promoted grain recruitment and broader stress distribution, leading to reduced transmitted peak forces and enhanced energy dissipation. Introducing elliptical voids in the architected grains further amplified these effects. The compliant internal architecture enables greater lateral deformation during impact, which increases penetration depth and impact duration while lowering the peak force. The combined outcome is a substantial enhancement in energy dissipation per unit mass. To isolate the influence of inclusion design parameters, such as aspect ratio and relative density, we systematically varied these features within the FEM framework and quantified their effects on both normalized kinetic energy ($T^*=T/T_0$, where $T_0$ is the initial kinetic energy of the impactor) and transmitted peak forces ($F^{*}_{max}=F_{max}/ER^{2}$) during impact---key metrics for protective performance---summarized through a unified impact mitigation parameter, $\eta = {T^*}/{\bar{\rho}F^{*}_{max}}$ (Fig.~\ref{fig:fig4_simulations}\textbf{b}). Simulations mirrored the trends observed in LIPIT experiments: reducing the relative density of the grains increased specific energy dissipation, primarily through enhanced frictional losses arising from greater inter-grain motion.

To expand the design space beyond idealized crystalline arrangements, we introduced programmable motifs inspired by crystallography---a third level of hierarchy in our granular metamaterials (Fig.~\ref{fig:fig4_simulations}\textbf{c}). These non-periodic configurations incorporate spatial variations to locally tailor the mechanical response. Specifically, we explored four representative motifs: (i) interstitials, formed by inserting smaller grains between adjacent ones to enhance local contact networks; (ii) fill gradients, created by varying the relative density of grains to combine structurally rigid and energy-dissipative regions; (iii) size gradients, generated by systematically varying grain diameters to promote force delocalization through diverging stress chains; and (iv) vacancies, introduced by selectively removing grains to induce greater rearrangement and frictional dissipation during impact. Finite element simulations of these spatially varied packings revealed that all configurations exhibited enhanced energy dissipation compared to uniform packings (Fig.~\ref{fig:fig4_simulations}\textbf{d}). Namely, the interstitial motif increased grain recruitment through the formation of additional vertical force chains, while the fill and size gradients promoted heterogeneous stress redistribution across layers and enhanced delocalization. Vacancies, on the other hand, induced greater particle mobility, amplifying frictional energy loss. The enhanced energy dissipation achieved by incorporating vacancies within the architected granular packing was further validated via additional LIPIT experiments (Extended Data Fig. 7). Beyond improving energy dissipation, these defect-engineered packings also reduced peak transmitted forces during impact. Together, these results demonstrate that introducing spatial non-uniformity and hierarchical organization enables simultaneous enhancement of energy dissipation and reduction of peak impact forces—establishing a powerful strategy for designing lightweight, impact-mitigating granular metamaterials.

\subsection{Concept implementation at the macroscale}
\begin{figure}[H]
    \centering
    \includegraphics[width=1.0\textwidth]{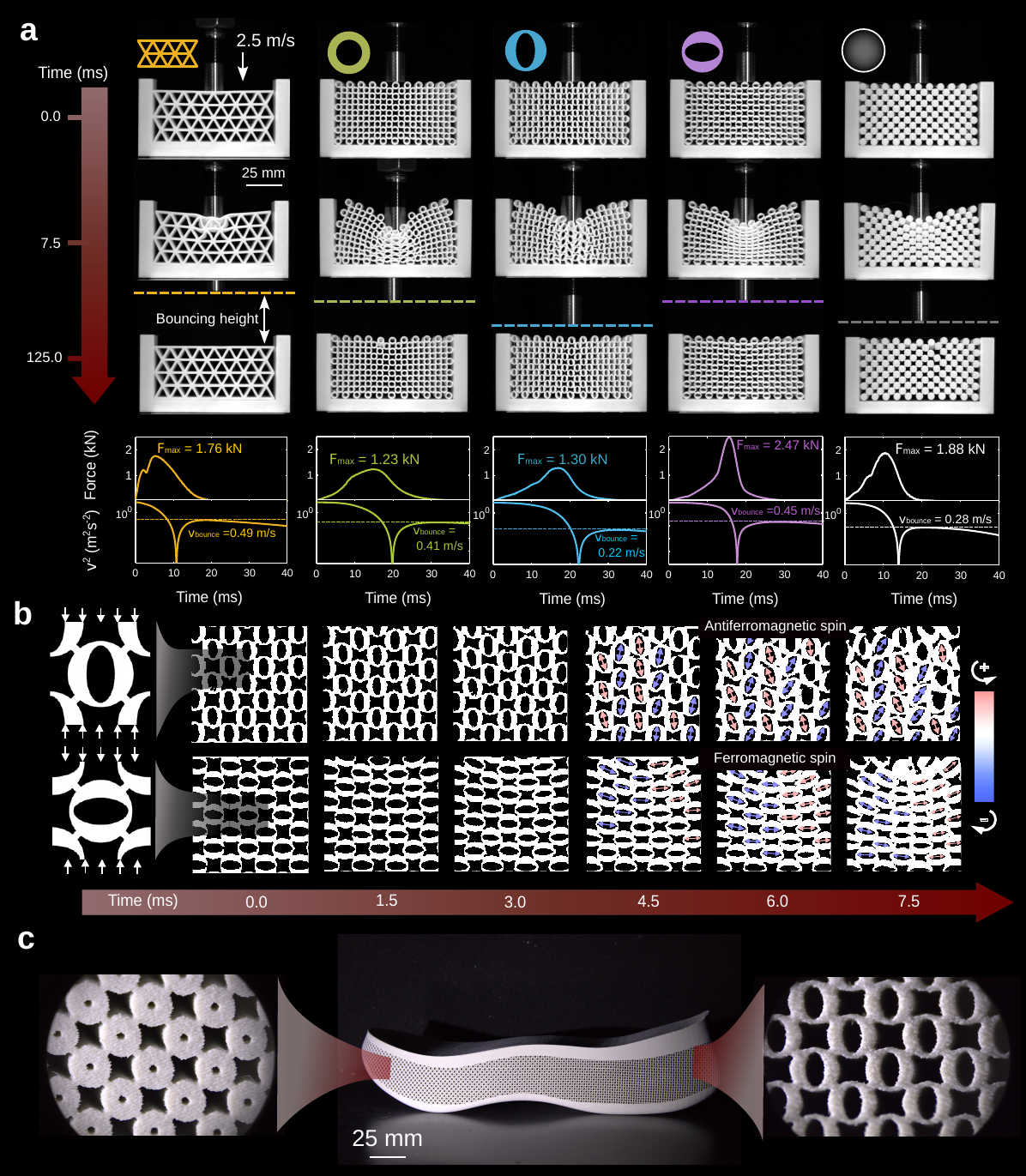}
    \caption{\textbf{Granular-inspired architectures at the macroscale for programmable energy dissipation and force mitigation.} \textbf{a}, Low-velocity impact response of granular-inspired architectures and comparison to a triangular lattice; with relative density of $\bar{\rho}$ = 46\% for the triangular lattice, 43\% for the architected grains, and 55\% for the monolithic grain design. \textbf{b} Local spin of architected grains during the impact event; the grains with vertical voids undergo antiferromagnetic spin, a phenomenon that facilitates the relative slipping among neighboring grains and leads to higher frictional energy dissipation. \textbf{c}, Granular-inspired sole featuring spatially programmed stiffness for customized energy dissipation and force mitigation, fabricated out of TEPU 50A elastomer via Vision-Controlled Jetting technology (Inkbit, Boston, MA, USA).}
    \label{fig:fig5_macro}
\end{figure}

\hfill\\
Building upon mechanistic insights from microscale experiments and the computational models, we extended the hierarchical concept to granular-inspired architectures at the macroscale, demonstrating scalability across length scales and material systems. We additively manufactured centimeter-scale 2D tessellations out of thermoplastic polyurethane (TPU), and evaluated their energy dissipation and force mitigation performance using drop tower impact experiments (Fig.~\ref{fig:fig5_macro}\textbf{a}).

Impact experiments on these macroscale samples revealed that architected grains significantly reduce peak transmitted forces and increase specific energy dissipation compared to conventional lattices of similar mass and geometry (Fig.~\ref{fig:fig5_macro}\textbf{a} and Extended Data Fig. 8). At the macroscale, the inclusion aspect ratio and orientation exert a more pronounced influence on the impact response than at the microscale. This sensitivity arises from two key factors: (i) the greater local deformability afforded by TPU, and (ii) thicker inter-grain connectors, which constrain global rearrangement while promoting local rotations and internal grain deformation. These conditions amplify the role of grain anisotropy in tuning the mechanical response. Among the tested configurations, grains with vertically oriented elliptical inclusions demonstrated the highest peak impact force minimization and energy dissipation. This performance stems from their enhanced capacity for antiferromagnetic spin motion\cite{coulais2018multi, coulais2018characteristic}, which increases inter-grain friction and localized deformation during compression (Fig.~\ref{fig:fig5_macro}\textbf{b}). In contrast, horizontal or circular inclusions favor axial compaction, limiting energy dissipation and resulting in higher transmitted forces. 

Leveraging the spatial programmability of the dynamic response, we translated the concept into functional prototypes, including architected granular midsoles (Fig.~\ref{fig:fig5_macro}\textbf{c}). By selectively varying grain architecture and arrangement, we tailored the regional mechanical response (Supplementary Note 6 and Extended Data Fig. 9). Zones with relatively high stiffness provide responsive propulsion (e.g., forefoot of a running shoe), and zones with high dissipation and low peak force provide comfort and impact protection (e.g., heel or orthopedic supports). These macroscale implementations of hierarchical granular metamaterials demonstrate the applicability of this framework for scalable, tunable, and reusable energy-absorbing systems that bridge mechanical functionality with user-centric applications.

\subsection{Expanding the design space: programmable contact as a multifunctional platform}
\begin{figure}[H]
    \centering
    \includegraphics[width=1.00\textwidth]{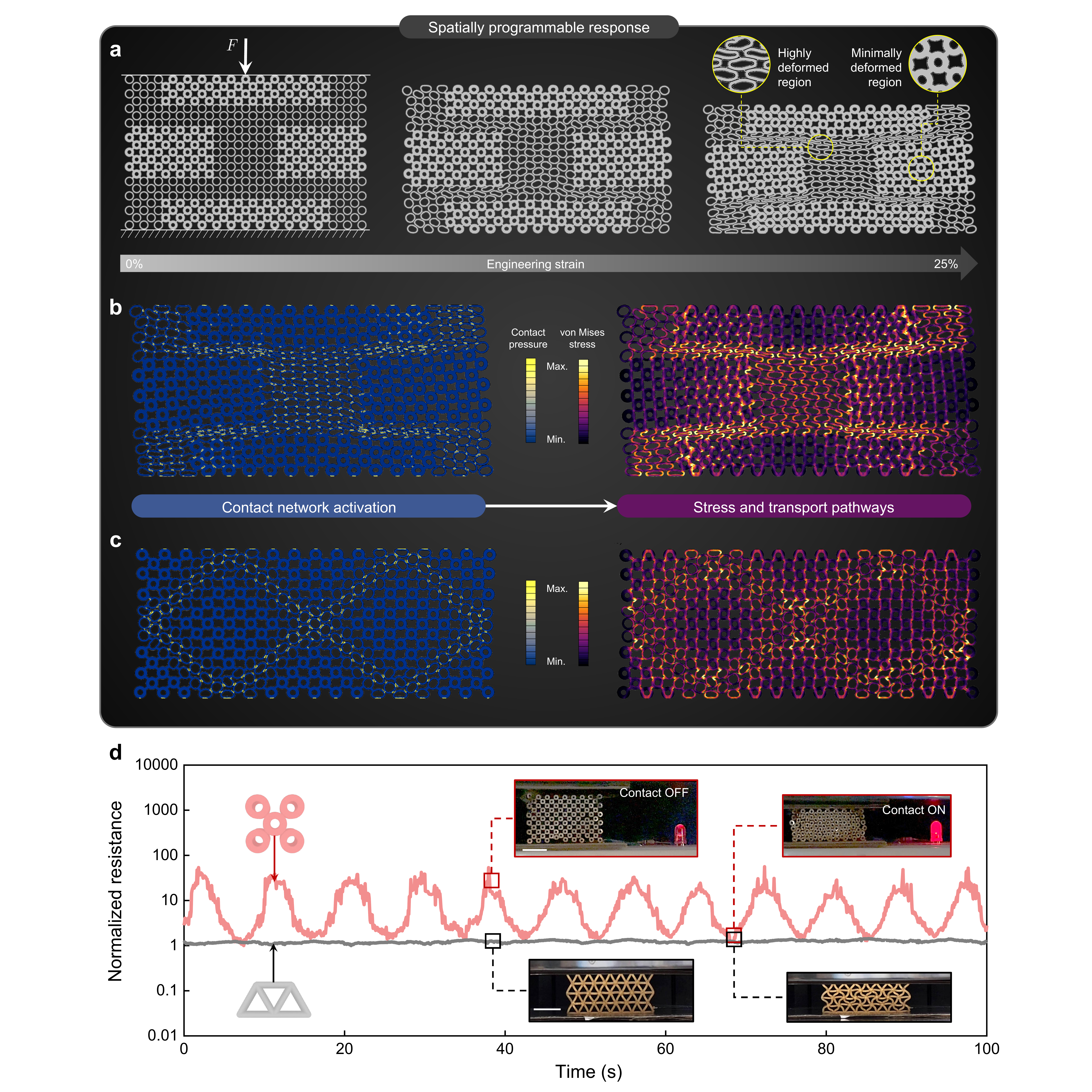}
    \caption{\textbf{Spatially programmable multifunctional response in heterogeneous granular metamaterials.} \textbf{a}, Simulated deformation sequence of a metamaterial with spatially graded unit-cell geometry under uniaxial compression. The heterogeneous architecture directs buckling into discrete highly deformed and minimally deformed regions, as depicted in the magnified insets. \textbf{b}, Finite-element models showing the evolution from contact-pressure distribution to von Mises stress and transport pathways, illustrating how local contacts collectively establish long-range contact networks. \textbf{c}, Corresponding simulations for an alternative heterogenous granular metamaterial design, demonstrating that the spatial patterning of contact network and resulting stress pathways can be independently programmed through geometry. \textbf{d}, Experimental validation of contact-activated electrical switching. Cyclic compression of the architected granular sample reproducibly induce resistance variation greater than one order of magnitude upon contact activation, as captured by the representative photographs of the uncompressed (Contact OFF) and compressed (Contact ON) device states, whereas little to no change in resistance observed in the triangular lattice sample of comparable mass and size due to the absence of self-contact. Scale bar, 1 cm.}
    \label{fig:fig6_multifunctional}
\end{figure}

\hfill\\
In conventional granular packings and mechanical metamaterials, contact networks and force chains are emergent phenomena determined entirely by packing geometry and applied boundary conditions, and therefore not independently controllable. Here, we demonstrate that by encoding spatially varying connecting struts directly into the grain architecture, contact pathways can be prescribed \textit{by design}, decoupled from packing configuration and independently programmed across the material volume (Fig.~\ref{fig:fig6_multifunctional}, Extended Data Fig. 10, Supplementary Note 8). This introduces a previously inaccessible degree of freedom in hierarchical granular metamaterials: the ability to spatially localize and route inter-grain contact---and by extension, stress, electrical, and thermal transport---as an intrinsic architectural property rather than a boundary condition or packing-dependent outcome.

Under uniaxial compression, the heterogeneous granular metamaterial directs buckling selectively into predefined regions, producing discrete heterogeneous deformation zones that would not arise from packing geometry alone (Fig.~\ref{fig:fig6_multifunctional}\textbf{a}). This spatially programmable deformation response is captured in finite-element simulations that reveal two coupled stages: first, the selective activation of a contact network within prescribed regions, visualized through inter-grain contact pressure distributions; and second, the emergence of long-range stress and transport pathways that are channeled along the geometry-encoded contact network (Fig.~\ref{fig:fig6_multifunctional}\textbf{b}). Critically, the topology of these pathways is not set by loading conditions but is written directly into the metamaterial architecture---as confirmed by an alternative grain design that programs an entirely different contact and stress-transmission pattern under identical boundary conditions (Fig.~\ref{fig:fig6_multifunctional}\textbf{c}). Together, these results demonstrate that contact networks can be independently engineered across the material volume by tuning local strut geometry, substantially expanding the design space beyond what packing configuration alone can achieve.

The functional consequences of this architecture-driven contact control extend directly into the electrical domain. Because engineered inter-grain connections govern interfacial contact area and connectivity, compression-induced contact percolation triggers a sharp, geometry-mediated piezoresistive response. When integrated into a simple circuit with a light-emitting diode (LED), the metamaterial functions as a mechanically gated switch: cyclic compression reproducibly modulates the resistance by over an order of magnitude across repeated loading cycles, whereas classical triangular lattices exhibit negligible resistance change under comparable strains due to the absence of evolving contact networks (Fig.~\ref{fig:fig6_multifunctional}\textbf{d}).
Beyond electrical transport, the same programmable contact topology can potentially be extended to thermal conductivity, where load-dependent contact networks selectively open or suppress thermal transport pathways. This concept could enable mechanically reconfigurable thermal transport, including directed dissipation from localized hotspots or confinement of thermal flux to prescribed regions, offering a route toward lightweight, mechanically adaptive thermal management for high-power electronics.

Overall, this framework positions programmable contact as a versatile design strategy for coupling mechanical stimuli to functional responses across multiple physical domains. By embedding functionality within deformation-governed contact networks, hierarchical granular metamaterials can be designed to exhibit load-responsive behavior in which structure, mechanics, and transport are closely linked. While the present results focus on electrical switching via piezoresistivity, the same principles suggest broader opportunities for mechanically mediated control of other transport phenomena.

\subsection{Summary and outlook}
\hfill\\
In this work, we introduced a class of hierarchical granular metamaterials that leverage architected grains, programmable packing configurations, and engineered inter-grain connections to achieve tunable mechanical and functional responses across length scales. By combining computational modeling with laser-induced particle impact testing and drop tower experiments, we systematically revealed how structural features at three hierarchical levels—grain architecture, packing topology, and spatial heterogeneity—govern quasi-static and dynamic mechanical behavior, as well as emergent transport responses.These results show that discretizing a lightweight architecture into compliant, interacting grains shifts impact mitigation from localized strut collapse to a collective, contact-mediated process, enabling lower transmitted peak forces together with higher mass-normalized energy dissipation.

Our results show that deformable architected grains containing hollow, anisotropic inclusions undergo enhanced lateral expansion, promoting the recruitment of neighboring grains and increasing dissipative plastic and frictional interactions during impact. Inspired by crystallographic defects and natural toughening strategies, we introduced spatial heterogeneity through functional gradients, vacancies, and interstitial-like motifs, thereby perturbing force-chain networks and enhancing energy dissipation while reducing transmitted peak forces. These hierarchical arrangements outperform conventional lattice materials at comparable densities, highlighting how discrete grain interactions introduce dissipation mechanisms beyond strut bending or stretching alone. In doing so, this framework bridges the discrete, contact-mediated mechanics that have long defined granular media\cite{liu1995force} with the deterministic design control of architected materials, providing a route to systematically program the force-chain topologies that classical granular systems can only access stochastically.

Beyond impact mitigation, we demonstrate that programmable contact pathways between architected grains enable controlled modulation of electrical transport. By engineering tunable compliance and controlled buckling within spatially varying connecting struts, inter-grain contact networks can be selectively activated under compression, leading to a pronounced and reversible electrical response. While demonstrated here for electrical switching, the same contact-mediated mechanism is expected to influence other transport processes, such as thermal conduction, suggesting broader opportunities for coupling mechanical deformation to functional behavior. 

More broadly, this framework draws inspiration from crystallography and metallurgy to inform the design of granular architected systems, in which grain-level building blocks act as tunable analogues to atoms. Unlike atomic crystals, however, these systems enable the deliberate spatial programming of defects, force-chain topology, and contact-mediated interactions, providing access to mechanical responses and coupled behaviors that are challenging to achieve in monolithic solids. Importantly, these design principles are demonstrated across both microscale and macroscale architectures and different polymeric systems, suggesting their applicability across length scales and material platforms. Notably, a wide range of effective responses is already accessed within polymeric constituents by spanning stiff photoresist microarchitectures and compliant TPU macroscale systems, indicating that the design space is not tied to a single material behavior.

Taken together, these results point toward a broader design strategy in which hierarchical granular metamaterials can be engineered as load-responsive systems with tunable mechanical and transport behavior. Future efforts may focus on integrating stimuli-responsive materials, extending deformation mechanisms to include plasticity and fracture, and developing inverse design approaches for application-specific architectures. In this context, the discrete, contact-mediated interactions that have historically complicated the analysis of granular materials may instead provide a foundation for their systematic design.

\subsection{Methods}
\hfill\\
\noindent\emph{Fabrication of hierarchical granular metamaterials}
\hfill\\
We fabricated monolithic microscale granular packings using CAD models developed in SolidWorks (Dassault Systèmes SE). Each geometry was exported and prepared for two-photon polymerization (TPP) printing on the Nanoscribe Photonic Professional GT2 system. Post-printing, the granular samples were developed in propylene glycol methyl ether acetate (PGMEA) for 24 hours and rinsed in isopropanol (IPA) for 15 minutes. Finally, they were dried using a critical point dryer to avoid capillary collapse. These packings were printed inside cylindrical “granular wells” (outer diameter: 200 \textmu{}m, inner diameter: 144 \textmu{}m, depth: 100 \textmu{}m), which served as a confined boundary during mechanical characterization. To further minimize the mechanical influence of the connector struts while still preserving structural integrity before compression, the packings underwent oxygen plasma ashing, selectively thinning the struts. This treatment allowed the granular wells to retain structural integrity in the undeformed state while enabling mechanical behavior representative of an ideal unconstrained granular system. 

\noindent\emph{In situ quasi-static mechanical characterization}
\hfill\\
\emph{In situ} quasi-static indentation experiments were performed on the fabricated microscale granular packings within a scanning electron microscope (SEM, Gemini 450, ZEISS) using an Alemnis Nanoindenter (Alemnis AG) equipped with a 50 \textmu{}m flat punch tip. Indentation experiments were conducted under displacement control at a nominal strain rate of $1 \times 10^{-2} \, \text{s}^{-1}$, enabling direct visualization of grain-level deformation and rearrangement mechanisms during loading. The stress-strain data was derived by normalizing the load-displacement data with the nominal area and height of each specimen, respectively. The dissipated energy density for each sample was obtained by integrating the area under the stress-strain curve.  

\noindent\emph{Dynamic mechanical characterization}
\hfill\\
To experimentally validate the dynamic impact behavior of architected granular packings at the microscale, we employed laser-induced projectile impact testing (LIPIT) whose setup is shown in Fig.~\ref{fig:fig3_packing}\textbf{g}. In this method, individual microparticles were launched from a prepared substrate toward a target and monitored in real-time with high-speed imaging. The launch substrate consisted of a 210 \textmu{}m-thick, 25 mm-diameter glass cover slip coated with a 60 nm sacrificial gold film (via sputter deposition) and a 30 \textmu{}m polyurea layer (applied by spin coating). Prior to testing, 50 \textmu{}m silica microparticles were placed onto the polyurea surface, evenly dispersed using ethanol and lens-cleaning paper. Immediately before firing, a single microparticle was chosen for acceleration. A pulsed Nd:YAG laser (532 nm wavelength, 10 ns pulse duration) was focused on the gold layer beneath the selected particle, causing localized ablation of the gold, rapid expansion of the underlying polyurea, and subsequent launch of the particle. By adjusting the laser pulse energy, the projectile velocity could be precisely controlled. The flight path and target zone were illuminated with a secondary laser (Cavilux, Specialized Imaging; 640 nm wavelength, 30 \textmu{}s pulse duration) and recorded through a microscope objective using an ultra–high-speed SIMX16 camera system (Specialised Imaging) equipped with 16 independently triggered intensified CCD sensors. Sequential frames from the recordings were analyzed to determine both impact and rebound velocities, as well as post-impact dynamics. Analysis of the response revealed that majority of the projectile’s kinetic energy is transferred into rearrangement of the grains and its motion. 

To quantify the energy dissipation and force mitigation characteristics of our macroscopic granular-inspired architectures, we performed low-velocity impact experiments using a drop-tower setup (Instron Dynatup 9250, Norwood, USA). A steel projectile was released at an impact velocity of $2.5 \  \mathrm{m} \mathrm{s}^{-1}$; the impact force and projectile velocity were recorded at a sampling frequency of 81.9 kHz over a duration of 99 ms. The impact event was captured utilizing a high-speed camera (Photron FASTCAM Nova S6, Tokyo, Japan) at 4000 frames $\mathrm{s}^{-1}$.

\noindent\emph{Finite element modeling}
\hfill\\
To complement experimental studies, we developed computational models capturing the mechanics of monolithic and architected granular systems across quasi-static and dynamic regimes. Implicit quasi-static indentation simulations in ABAQUS/Standard (Dassault Systèmes SE) on 2D granular packings confined within rigid boundaries. The constitutive model consisted of an elastoplastic response with isotropic strain-hardening, calibrated to reflect the mechanical behavior of the IP-Dip photoresist, obtained through micropillar compression experiments (Supplementary Fig. 1). The calibrated properties consisted of Poisson’s ratio $\nu = 0.36$, density $\rho=1200$ kg/m\textsuperscript{3}, and Young’s modulus $E= 3.1$ GPa. Each grain geometry was meshed with CPE4R elements, using ~200 elements along the periphery and a minimum of 5 elements spanning the thickness. Contact behavior was modeled using general contact with Coulomb friction, adopting an isotropic friction coefficient of 0.4 (a commonly used value for polymer-on-polymer contact). Quasi-static loading was applied through a rigid indenter displaced to 50\% engineering strain, while an encastre boundary condition was imposed on the confinement. 

To probe dynamic response, the framework was developed further to include implicit  and explicit dynamic simulations for 2D and 3D packings, respectively, in ABAQUS/Standard (Dassault Systèmes SE). A rigid projectile prescribed with an initial velocity was allowed to impact the granular assembly, while maintaining the same meshing parameters and encastre boundary condition on the confinement. The transmitted force at the base and the projectile’s kinetic energy was extracted to characterize the force mitigation and energy dissipation properties. 

\subsection{Data availability}
\hfill\\
The data that support the findings of this study are present in the paper and/or in the Supplementary Information. Additional data related to the paper are available from the corresponding author upon request.

\subsection{Acknowledgements}
\hfill\\
This work was performed in part in the MIT.nano Fabrication and Characterization Facilities. Research was sponsored by the Army Research Office and was accomplished under Cooperative Agreement Number W911NF-24-2-0182. The views and conclusions contained in this document are those of the authors and should not be interpreted as representing the official policies, either expressed or implied, of the Army Research Office or the U.S. Government. The U.S. Government is authorized to reproduce and distribute reprints for Government purposes notwithstanding any copyright notation herein. The authors thank S. Figueroa for his assistance in developing the initial designs and realizations of architected grain samples; and J. Ramos, J. René, and D. Chen from Inkbit for contributing elastomeric realizations of our metamaterials; and A.Y. Chen for assistance with X-ray tomography. 

\subsection{Author contributions}
\hfill\\
C.M.P., B.F.G.A., and J.U.S. conceived this study. J.U.S. designed the architectures, fabricated microscale samples, performed mechanical experiments and analyzed data. A.K, B.F.G.A. and J.U.S. performed simulations and analyzed data. J.L. and A.K. performed LIPIT experiments and analyzed data. L.W. designed, fabricated and tested macroscale samples and analyzed data. K.K. and B.F.G.A. wrote and validated the theory. J.U.S. and L.W. fabricated conductive macroscale samples and performed  electromechanical experiments. C.M.P. supervised the project. J.U.S, B.F.G.A and C.M.P. wrote the manuscript with input from all authors.

\subsection{Competing interests}
\hfill\\
The authors have filed a patent for the hierarchical granular metamaterials presented in this work.

\nolinenumbers
\vspace{10pt}

\subsection{References}
{\spacing{1}
\bibliography{references}
}
\bibliographystyle{naturemag.bst}

\setcounter{figure}{0}
\newpage
\subsection{Extended data figures and tables}
\hfill\\
\begin{figure}[H]
    \captionsetup{labelformat=extended,labelsep=vline}
    \centering
    \includegraphics[width=1.0\textwidth]{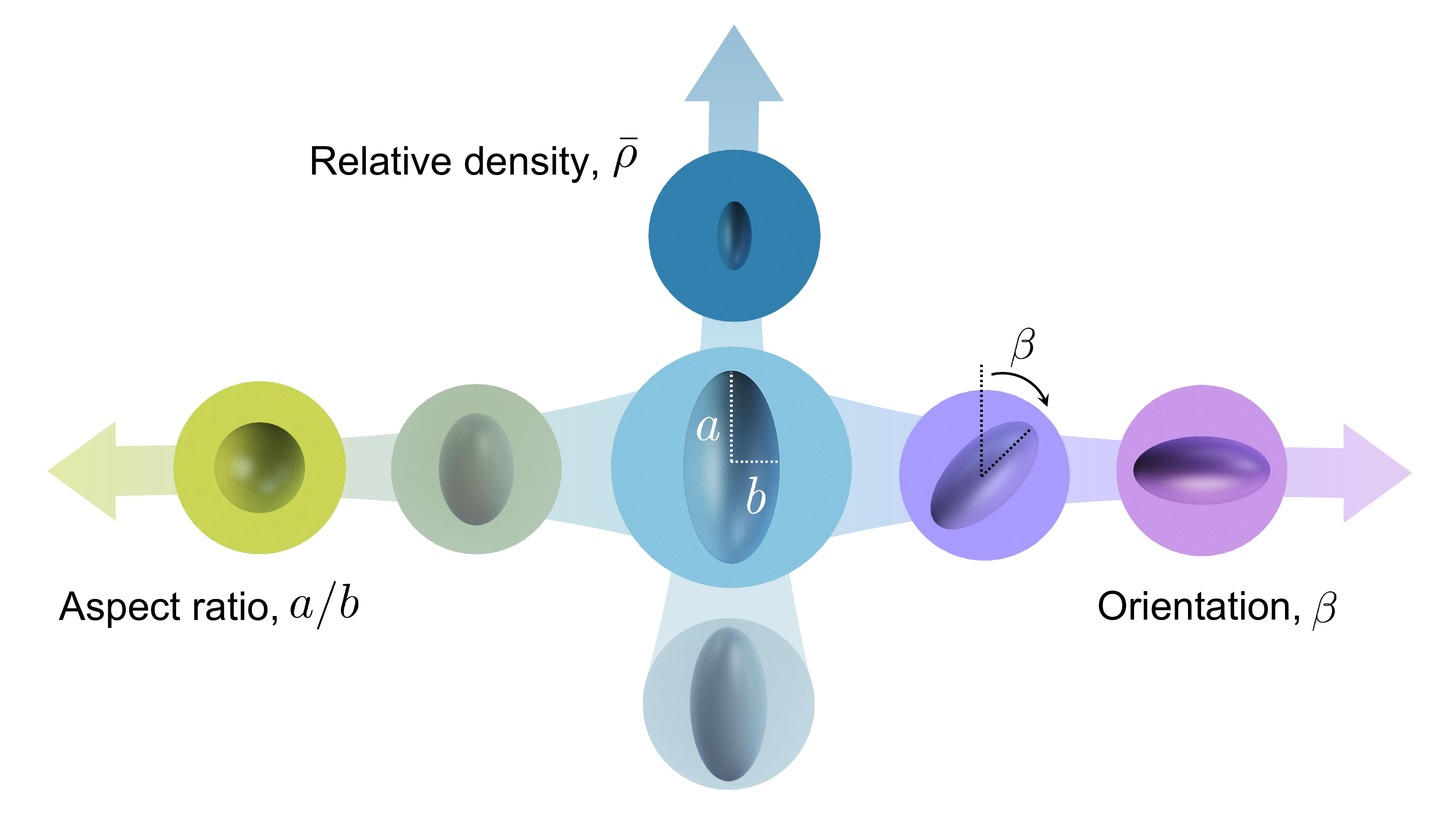}
    \caption{\textbf{Grain-level design space for hierarchical granular metamaterials.}
    Geometric design parameters defining architected grains with hollow elliptical inclusions. Shown are variations
    in inclusion aspect ratio ($a/b$), orientation ($\beta$), and relative density ($\bar{\rho}$), which collectively
    control anisotropic stiffness and deformation under load. These parameters form the first hierarchical level
    governing grain-scale mechanical response.}
    \label{fig:ed1_grain_design}
\end{figure}

\newpage
\begin{figure}
    \captionsetup{labelformat=extended,labelsep=vline}
    \centering
    \includegraphics[width=0.9\textwidth]{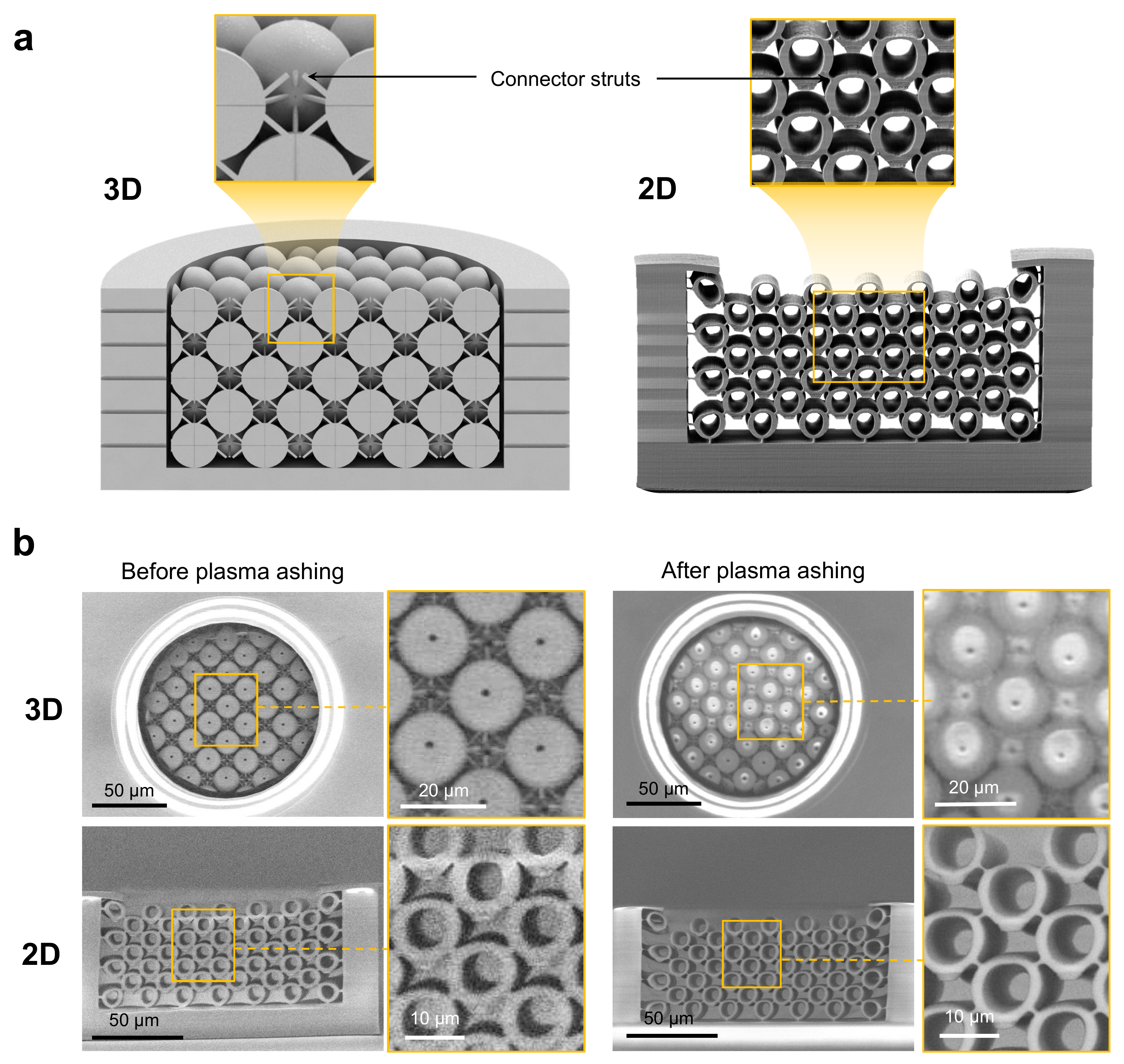}
    \caption{\textbf{Fabrication and confinement strategy for ordered granular packings.}
    \textbf{a}, Sacrificial inter-grain connectors used to preserve initial packing geometry prior to loading. \textbf{b} SEM images of fabricated 2D and 3D granular packings before and after connector removal.}
    \label{fig:ed2_connectors}
\end{figure}

\newpage
\begin{figure}
    \captionsetup{labelformat=extended,labelsep=vline}
    \centering
    \includegraphics[width=0.9\textwidth]{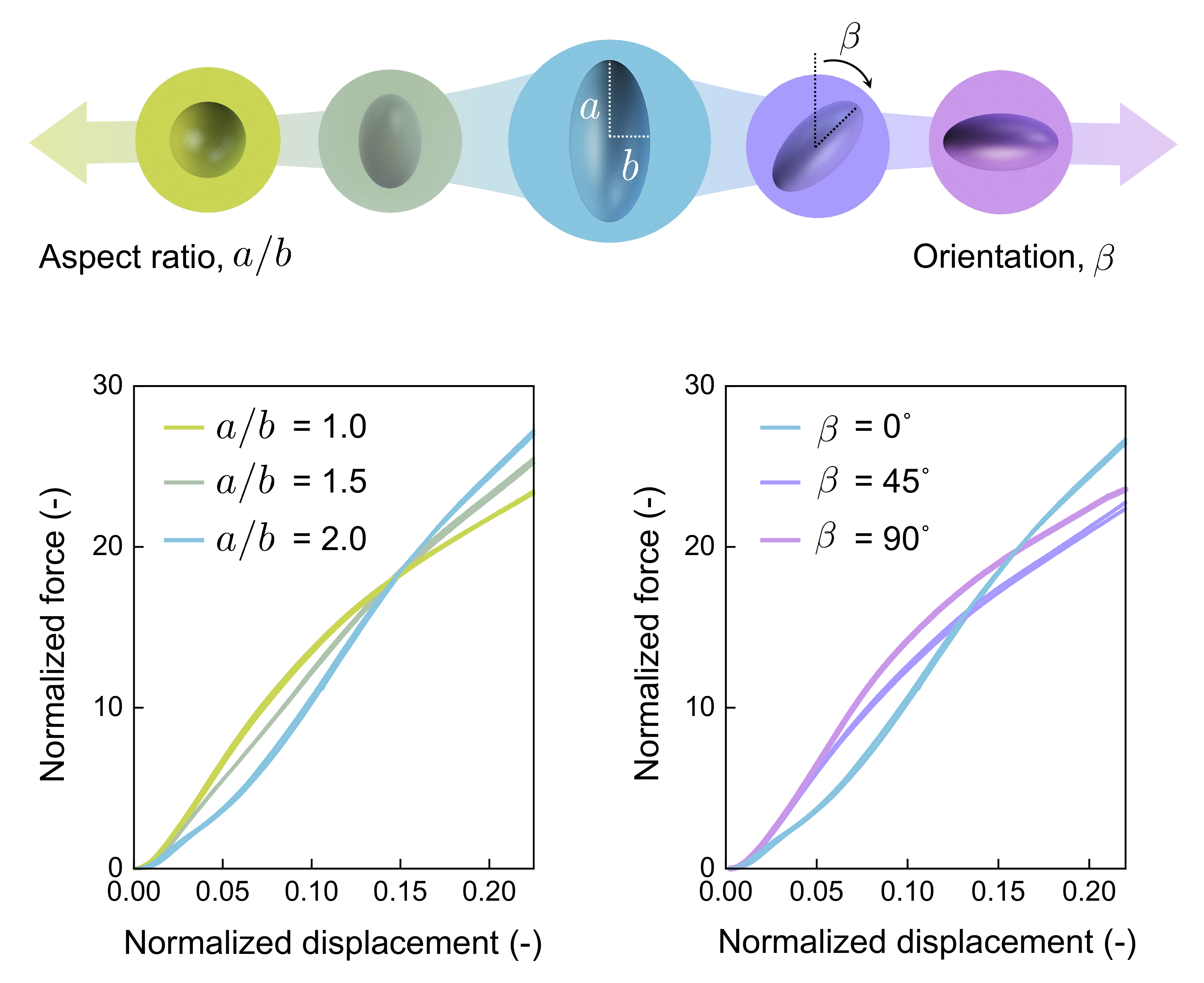}
    \caption{\textbf{Effect of inclusion geometry on single-grain mechanical response.}
    Engineering stress–strain response of architected grains with varying inclusion aspect ratios and orientations at fixed relative density.}
    \label{fig:ed3_inclusion_effect}
\end{figure}

\newpage
\begin{figure}
    \captionsetup{labelformat=extended,labelsep=vline}
    \centering
    \includegraphics[width=0.9\textwidth]{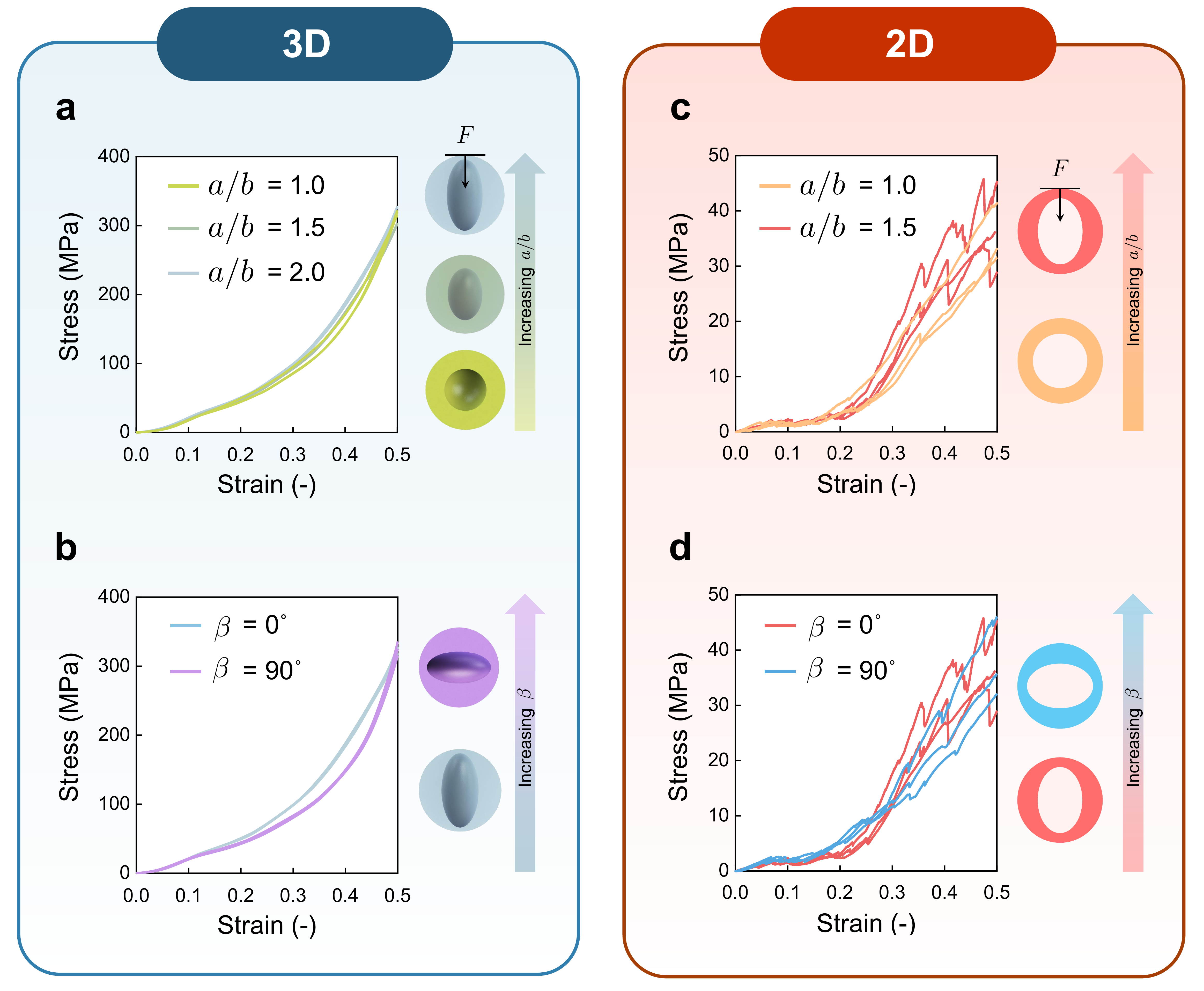}
    \caption{\textbf{Quasi-static indentation response of architected granular packings with varying inclusion aspect ratio and orientation.}
    Engineering stress–strain curves of 3D architected granular packings with various inclusion \textbf{a}, aspect ratios and \textbf{b} orientations at fixed relative density. Engineering stress–strain curves of 2D architected granular packings with various inclusion \textbf{c}, aspect ratios and \textbf{d} orientations at fixed relative density.}
    \label{fig:ed1_grain_design}
\end{figure}

\clearpage
\begin{figure}
    \captionsetup{labelformat=extended,labelsep=vline}
    \centering
    \includegraphics[width=0.95\textwidth]{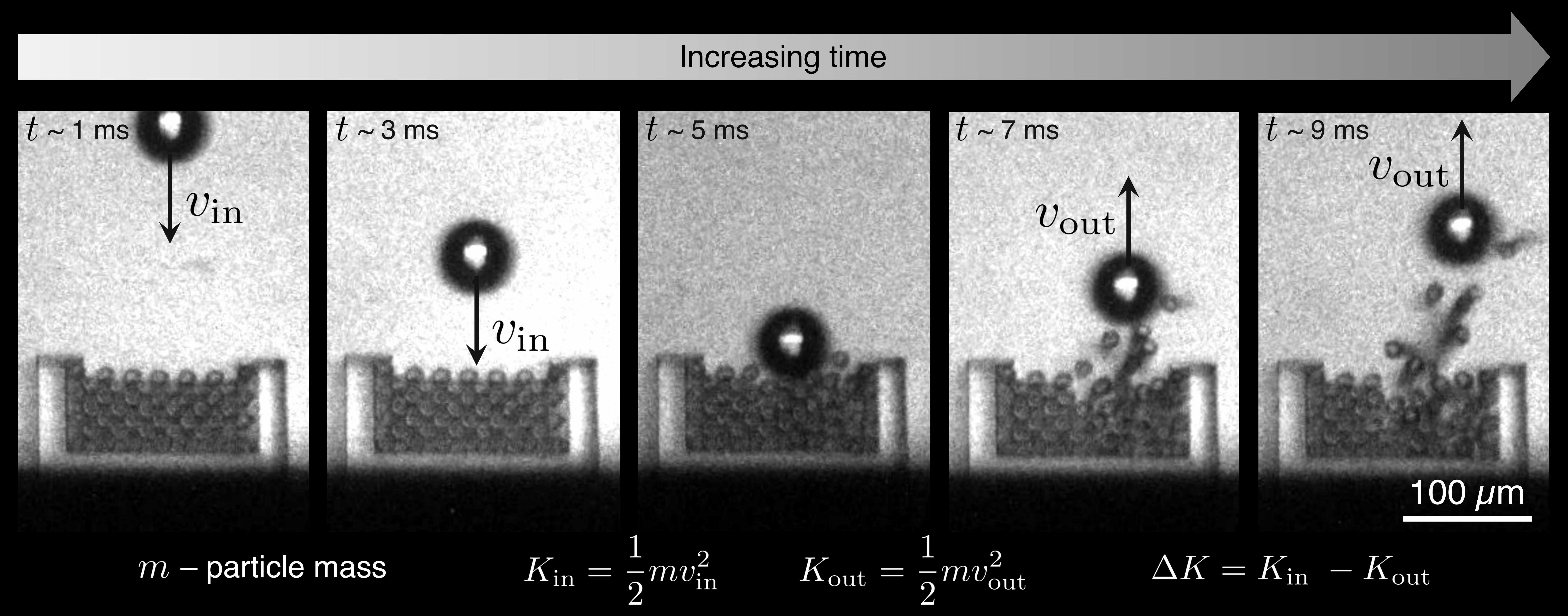}
    \caption{\textbf{Extraction of impact energy dissipation in LIPIT experiments.}
    \textbf{a}, Representative particle trajectories before and after impact on architected granular packings. \textbf{b} Method used to compute incident and rebound velocities from high-speed imaging. \textbf{c} Corresponding kinetic energy loss used to calculate normalized impact energy dissipation.} 
    \label{fig:ed1_grain_design}
\end{figure}

\clearpage
\begin{figure}
    \captionsetup{labelformat=extended,labelsep=vline}
    \centering
    \includegraphics[width=0.9\textwidth]{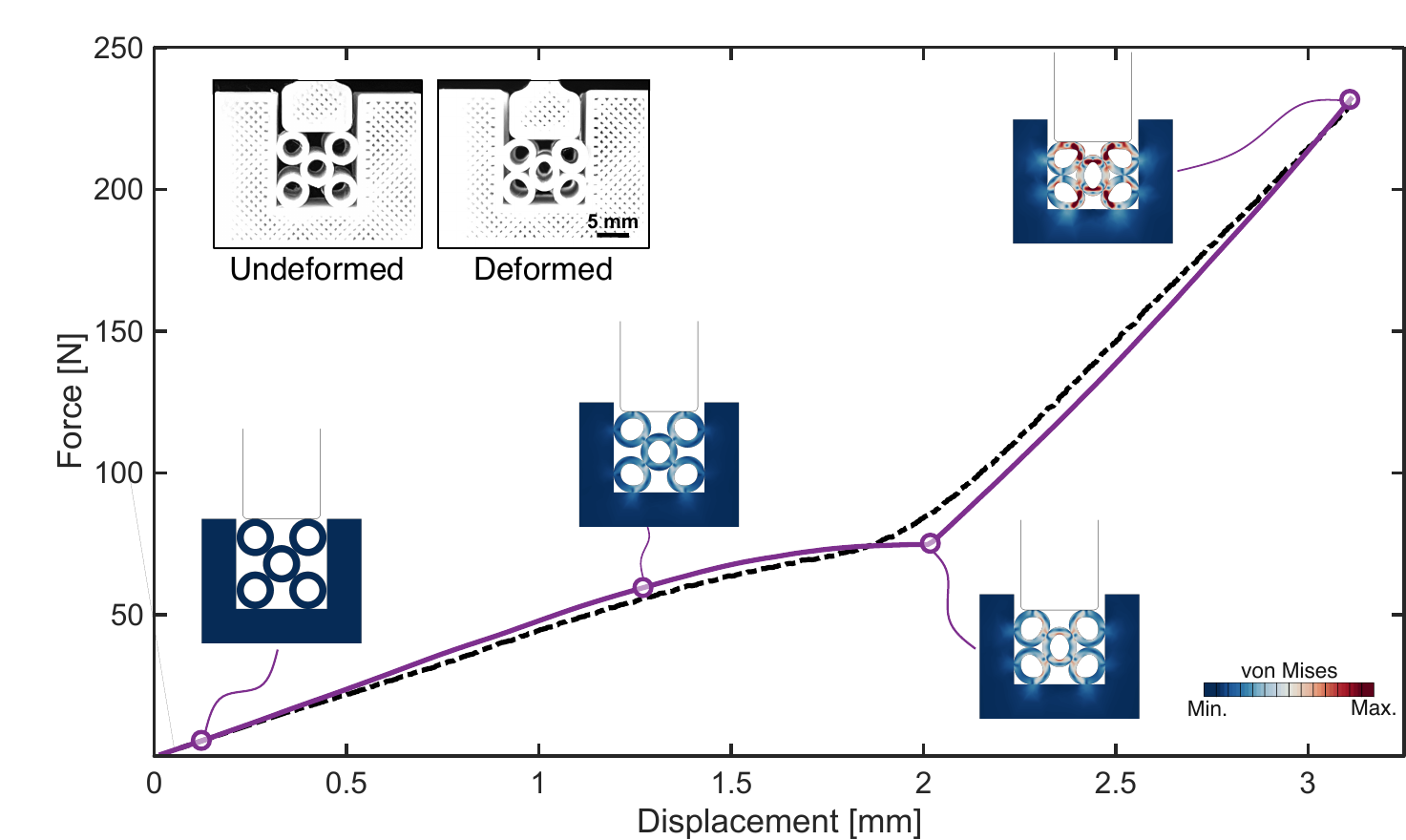}
    \caption{\textbf{Validation of FEM framework for 2D granular packings.}
    Comparison between experimental indentation results (black dashed curve) and finite element simulations (purple solid curve) for 2D architected granular packings. The model accurately captures force–displacement response and grain rearrangement patterns.}
    \label{fig:ed1_grain_design}
\end{figure}

\clearpage
 \begin{figure}
    \captionsetup{labelformat=extended,labelsep=vline}
    \centering
    \includegraphics[width=1.0\textwidth]{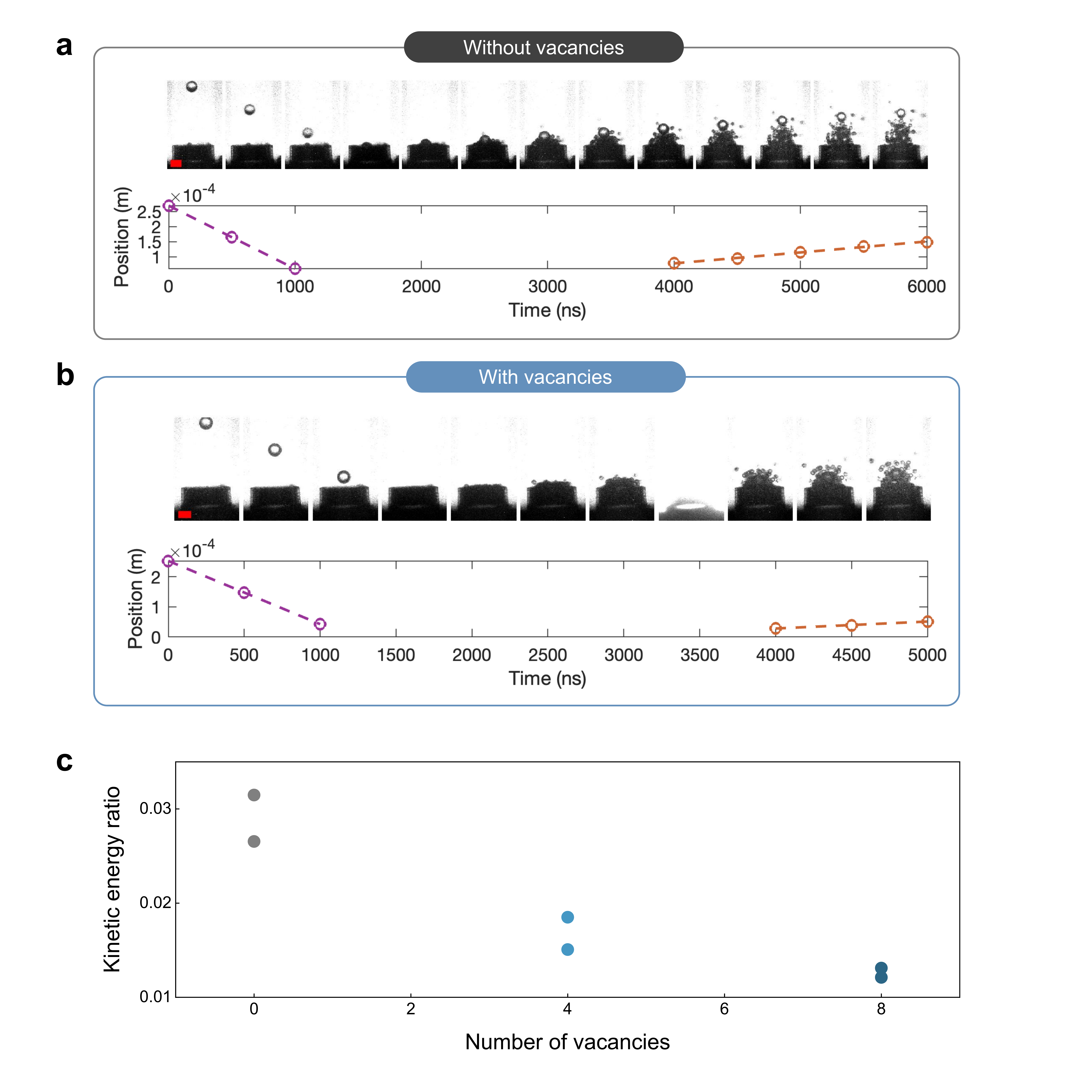}
    \caption{\textbf{Effect of vacancies on impact-induced energy dissipation.} \textbf{a, b}, Time-resolved snapshots and projectile position–time trajectories for impacts on granular assemblies \textbf{a} without vacancies and \textbf{b} with vacancies. In the vacancy-free assembly, the projectile rebounds after impact, reflecting elastic energy recovery. Introducing vacancies increases the available free volume for particle rearrangement and contact-network evolution, resulting in reduced rebound and enhanced dissipation of impact energy. Dashed lines indicate projectile trajectories before and after impact. Scale bars, 50 \textmu{}m. \textbf{c}, Rebound kinetic energy normalized by impact kinetic energy for assemblies containing different numbers of vacancies. The retained kinetic energy decreases with increasing vacancy number, indicating progressively greater energy dissipation enabled by vacancy-mediated structural reconfiguration.}
    \label{fig:metamaterial_cushion} 
\end{figure}

\clearpage
 \begin{figure}
    \captionsetup{labelformat=extended,labelsep=vline}
    \centering
    \includegraphics[width=\textwidth]{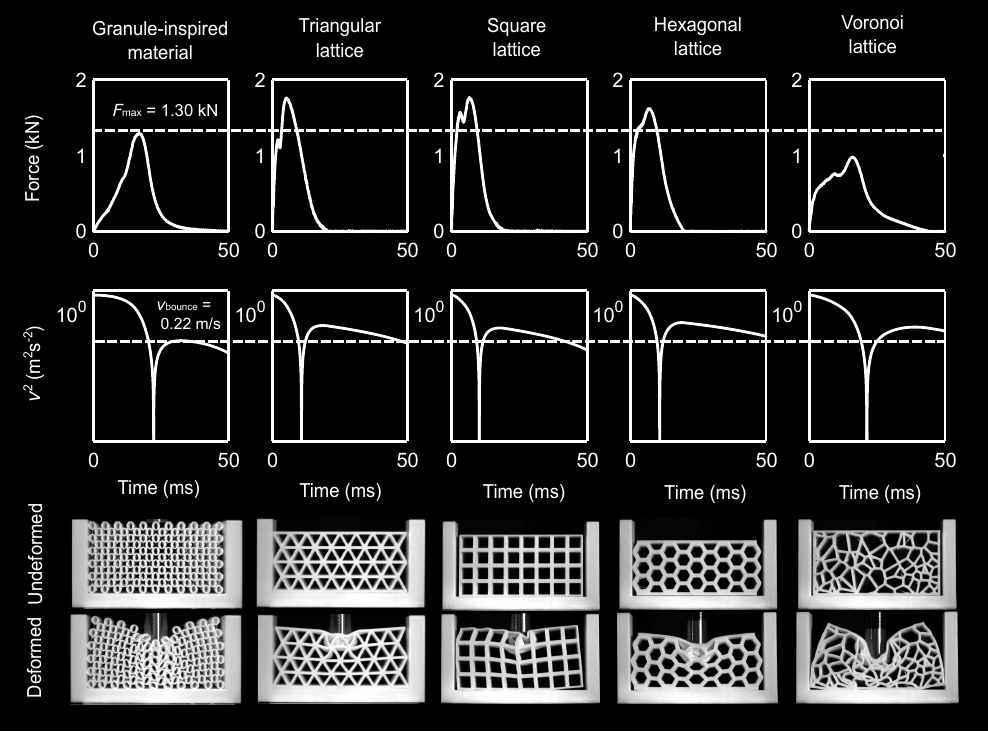}
    \caption{\textbf{Comparison between granular-inspired materials and conventional lattice materials.} }
    \label{fig:midsole} 
\end{figure}

\clearpage
 \begin{figure}
    \captionsetup{labelformat=extended,labelsep=vline}
    \centering
    \includegraphics[width=1.0\textwidth]{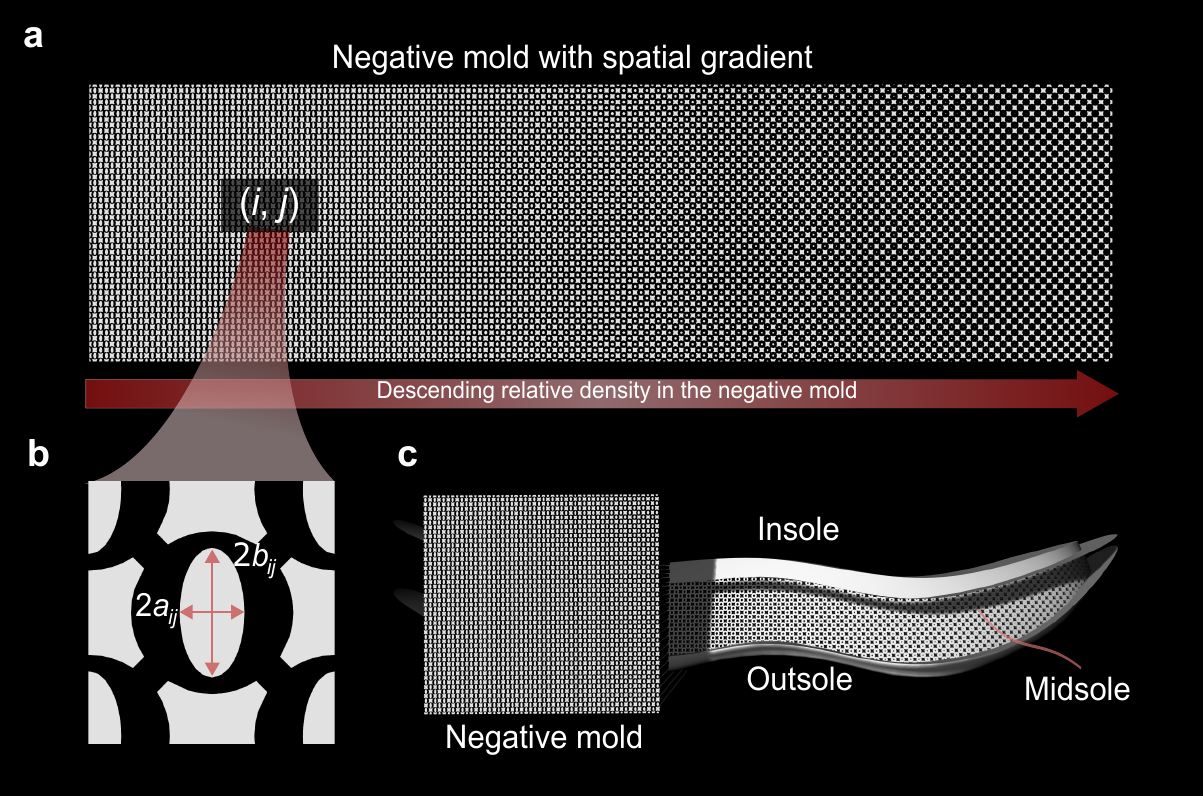}
    \caption{\textbf{Design of granular metamaterial midsole.} \textbf{a}, Negative mold with a monotonically descending relative density from left to right. \textbf{b}, Geometry of the elliptical inclusions as a function of the representative volume element index $(i,j)$. \textbf{c}, Construction of the granular metamaterial midsole via Boolean operations.}
    \label{fig:granule_lattice_compare} 
\end{figure}

\clearpage
\begin{figure}
    \captionsetup{labelformat=extended,labelsep=vline}
    \centering
    \includegraphics[width=0.9\textwidth]{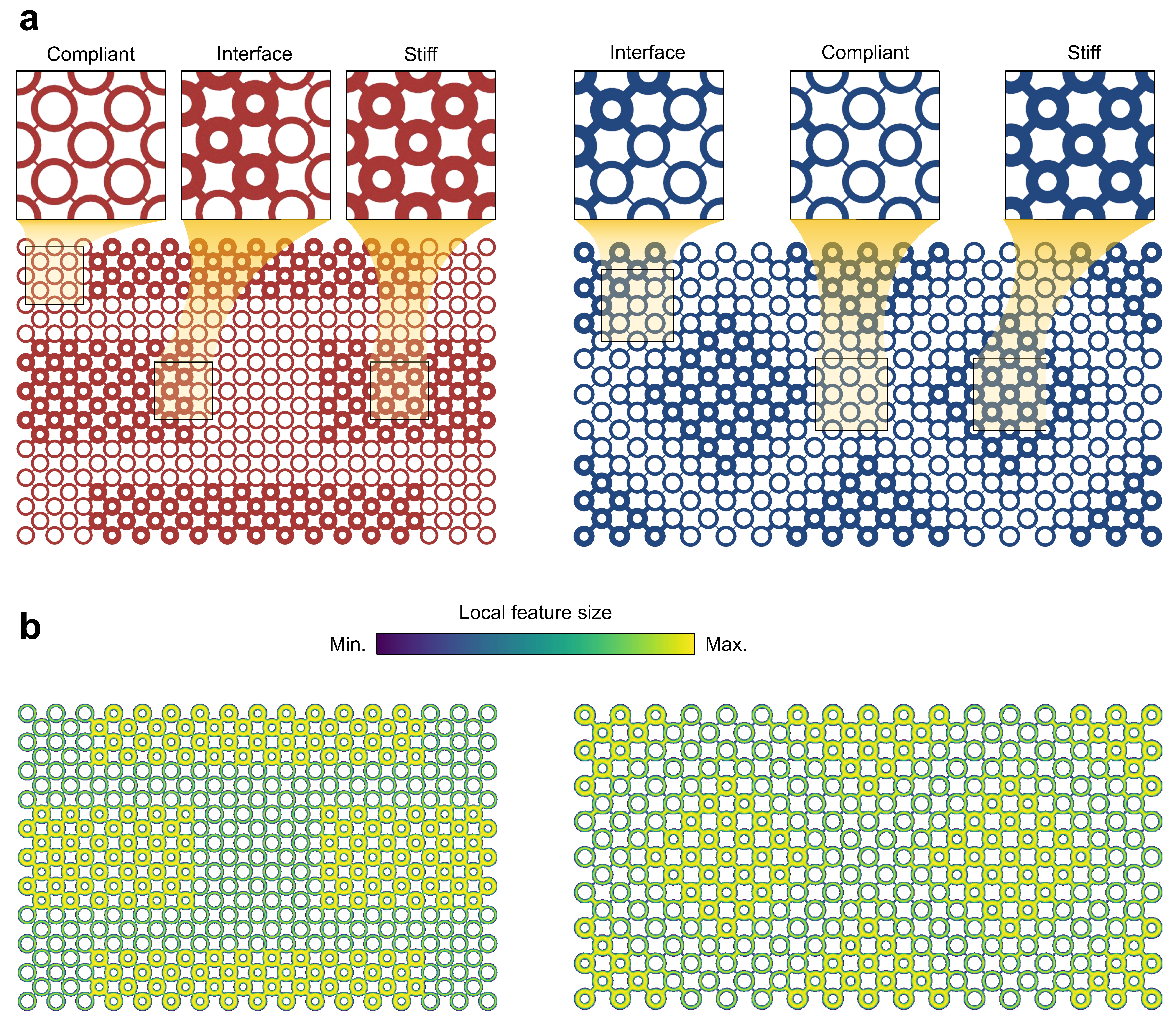}
    \caption{\textbf{Programming localized deformation in hierarchical granular metamaterials.} \textbf{a}, Granular architectures comprising compliant, interface, and stiff regions generated by spatially varying strut thickness. Thinner struts reduce local buckling resistance and promote deformation-induced contact activation, whereas thicker struts remain comparatively stable and suppress unintended contact formation. Insets show the corresponding local geometries. \textbf{b}, Maps of local feature size used to encode the spatial stiffness distribution. Controlled variation of strut thickness directs deformation along predefined pathways, enabling deterministic and reproducible contact-network formation under compression.}
    \label{fig:ed1_grain_design}
\end{figure}

\clearpage

\newcommand{\sigb}{\boldsymbol{\sigma}}

\newcommand{\epsb}{\boldsymbol{\epsilon}}

\setcounter{figure}{0}
\section{Supplementary information}

\subsection{Supplementary Note 1: Design of microscale architected grains}
\hfill\\
The microscale architected grains were designed as hollow spherical particles containing a centrally embedded elliptical inclusion (void). As illustrated in Fig.~S1, the mechanical response of each grain is governed by three independent geometric parameters: relative density ($\bar{\rho}$), inclusion aspect ratio ($a/b$), and inclusion orientation ($\beta$). Throughout this study, the loading direction is defined along the vertical axis, from top to bottom. The relative density is defined as the ratio between the solid volume of the architected grain and the volume of an equivalent fully solid sphere of identical dimensions, which is primarily controlled by the thickness of the spherical shell and the size of the internal elliptical void. The inclusion geometry is defined by an ellipse with major axis $a$ and minor axis $b$. The aspect ratio $a/b$ determines the elongation of the internal void while maintaining constant outer grain diameter. When $a/b = 1$, the inclusion is spherical (axisymmetric), resulting in isotropic stress distributions. Increasing $a/b$ elongates the void along its major axis, introducing geometric anisotropy and enhancing directional compliance along that axis. The orientation angle $\beta$ is defined as the angle between the major axis of the elliptical inclusion and the vertical loading direction. When $\beta = 0^\circ$, the major axis is aligned with the compression direction (top-to-bottom), maximizing axial compliance and facilitating deformation along the elongated void.

\subsection{Supplementary Note 2: Design of microscale architected granular packings}
\hfill\\
To ensure reliable fabrication of the microscale architected grains (outer diameter $D = 20$~\textmu{}m), additional design features were incorporated to preserve structural integrity during two-photon polymerization and subsequent solvent development. At this length scale, unsupported or weakly anchored particles are susceptible to detachment, drift, or collapse due to capillary forces and fluid flow during development and drying. To prevent individual grains from ``floating'' or being washed away, slender connector struts with a diameter of approximately 1 \textmu{}m were introduced between nominally contacting grains (Extended Data Fig. 2a). These connector struts functioned as temporary mechanical anchors, preserving positional registration and spatial organization of the granular assembly throughout printing and solvent development. Their diameter ($\sim$~1~\textmu{}m) was carefully chosen to balance two competing requirements: (i) sufficient bending stiffness and interfacial adhesion to resist capillary forces, viscous drag, and drying-induced stresses during development, and (ii) a minimal cross-sectional area to avoid artificial stiffening, load bridging, or premature force transmission during subsequent mechanical testing. Accordingly, the struts were designed to serve primarily as fabrication supports rather than structural reinforcement elements. To further suppress their mechanical contribution, an additional plasma ashing step was performed after development to reduce the strut diameter (Extended Data Fig. 2b). This post-processing treatment weakened the connectors while preserving grain geometry, thereby minimizing their influence on the measured response of the architected granular packings and ensuring that the observed mechanics were dominated by inter-granular interactions rather than by residual fabrication supports. In addition to the connector struts, a controlled inter-grain spacing of approximately 2~\textmu{}m was introduced between neighboring grains. This deliberate gap ensured that adjacent particles remained mechanically discrete after polymerization and post-processing. Without this spacing, unintended fusion or partial sintering between neighboring shells could occur, effectively transforming the granular assembly into a monolithic porous structure. Such bonding would suppress inter-granular sliding, rotation, and force-chain reconfiguration—mechanisms that are central to energy dissipation in granular media. By combining slender connector struts with a small but finite inter-particle gap, we ensured both fabrication robustness and mechanical separability. This design strategy preserves the intrinsic deformation, frictional sliding, and rearrangement mechanisms characteristic of granular systems, while maintaining structural fidelity and yield during microscale fabrication.

\subsection{Supplementary Note 3: Theoretical framework for predicting the mechanical response of architected grains}
\hfill\\
 To understand the mechanical behavior of an individual architected grain, let us consider a linear elastic sphere of radius $R$ and stiffness tensor $\mathbb{C}$, containing a centered ellipsoidal inclusion with semi-axes $(a,b)$. The sphere is subjected to a number of forces $\mathbf{f}_{i}$ on its surface. We further assume these forces are normal to the sphere's surface. We decompose this configuration into (1) an inclusion-free case, where the boundary forces are present but the sphere has no inclusion, and (2) an `Eshelby correction', where the sphere is still fully dense, but there are no boundary forces and the inclusion region is subjected to an eigenstrain $\epsb^{T}$. Under an appropriate choice of $\epsb^{T}$, the superposition of (1) and (2) will produce the intended configuration and boundary conditions, complete with traction-free boundaries along the inclusion. 
 
The objective of the correction problem is therefore to compute the inclusion-induced perturbation to the displacement field on the outer
surface of the sphere. We denote this perturbation by
$\mathbf{u}^{\mathrm{corr}}(\mathbf{x})$, so that the displacement of the sphere with the inclusion is approximated as
$\mathbf{u}^{(0)}(\mathbf{x})+\mathbf{u}^{\mathrm{corr}}(\mathbf{x})$,
where $\mathbf{u}^{(0)}$ is the displacement of the fully dense sphere
under the same applied forces. Once $\mathbf{u}^{\mathrm{corr}}$ is
determined, the displacement correction at the $i$-th contact is obtained by evaluating this field at the contact location.

First, for the no-void case (1), let $\boldsymbol{\sigma}^{(0)}$ denote
the stress field in the fully dense sphere. The stress tensor that
enters the Eshelby correction is the negative of the average of this field
over the domain $\Omega_{\mathrm{inc}}$ that will be occupied by the inclusion,
\[
\tilde{\boldsymbol{\sigma}}
=
-\frac{1}{|\Omega_{\mathrm{inc}}|}
\int_{\Omega_{\mathrm{inc}}}
\boldsymbol{\sigma}^{(0)}(\boldsymbol{x})\,{\rm d}V .
\]

In the absence of the exact finite-sphere solution for a general contact
network, we approximate $\tilde{\boldsymbol{\sigma}}$ by superposing the
Hertz stress magnitudes associated with each normal contact. Let the $i$-th contact be centered at $R\boldsymbol{n}_i$, where $\boldsymbol{n}_i$ is the unit vector from the sphere center to the contact point, and let $f_i$ be the magnitude of the corresponding normal force. We denote the Hertz contact radius by $c_i$, and write the peak contact pressure as
\[
p_{0i}=\frac{3f_i}{2\pi c_i^2}.
\]
For a Hertz contact, the axial compressive stress magnitude at a distance $d$ beneath the contact center is $p_{0i}(1+d^2/c_i^2)^{-1}$. Evaluating this expression at the center of the sphere ($d=R$) and retaining the stress component along the contact normal gives the approximation

\begin{equation}
\tilde{\boldsymbol{\sigma}}
\approx
\sum_{i=1}^{N_c}
\frac{p_{0i}}{1+R^2/c_i^2}
\boldsymbol{n}_i\otimes\boldsymbol{n}_i .
\end{equation}

For example, for the classic case of two equal diametrically opposed contacts along the vertical direction, $c_i=c$, $p_{0i}=p_0$, $\boldsymbol{n}_1=\boldsymbol{e}_3$, and $\boldsymbol{n}_2=-\boldsymbol{e}_3$, so that
\[
\tilde{\boldsymbol{\sigma}}
\approx
\frac{2p_0}{1+R^2/c^2}
\boldsymbol{e}_3\otimes\boldsymbol{e}_3,
\qquad
p_0=\frac{3f}{2\pi c^2}.
\]
Since the stress field varies strongly near the contacts but is expected to be slowly varying near the center of the sphere, we further assume that this averaged stress is constant throughout the inclusion domain. Our objective will then be to determine which eigenstrain
$\epsb^T$ generates this uniform compensating stress, thereby canceling the average no-void stress and yielding traction-free boundaries along the inclusion after superposition. From Eshelby, we know that the stress inside the inclusion, $\sigma_{ij}^{I}$, can be obtained as

\begin{equation}
    \sigma^I_{ij}
    =
    C_{ijkl}
    \left(
    \epsilon^I_{kl}
    -
    \epsilon^T_{kl}
    \right)
    =
    \tilde{\sigma}_{ij}.
    \label{eq:eshelby1}
\end{equation}

where $\epsilon_{kl}^{I}$ is the total strain inside the inclusion. We also know that 

\begin{equation}
    \epsilon_{ij}^{I} = S_{ijkl}\epsilon_{kl}^{T},
    \label{eq:eshelby2}
\end{equation}

where $S_{ijkl}$ is the Eshelby tensor. By solving Eq.~\eqref{eq:eshelby1} for the eigenstrain and using Eq.~\eqref{eq:eshelby2}, we get that

\begin{equation}
    \epsb^{T}
    =
    [\mathbb{C}(\mathbf{S}-\mathbf{I})]^{-1}
    \tilde{\boldsymbol{\sigma}}
    =
    (\mathbf{S}-\mathbf{I})^{-1}
    \mathbb{S}:\tilde{\boldsymbol{\sigma}} .
    \label{eq:transformation_stress}
\end{equation}

where $\mathbb{S} = \mathbb{C}^{-1}$ is the (fourth-order)  compliance tensor, and $\mathbf{I}$ is the fourth-order identity tensor. 

Let us now assume the deformed shape of the sphere containing an inclusion is, to first order, an ellipsoid. This is equivalent to stating that the displacement at a point $\mathbf{x}$ on the surface of the sphere can be written as $\mathbf{u}^{\text{corr}}=\mathbf{E}\mathbf{x}$, for some constant tensor $\mathbf{E}$. We will solve problem (2) by Betti reciprocity, considering an auxiliary problem where a traction $\boldsymbol{\check{\sigma}}\mathbf{n}$ for arbitrary, constant symmetric tensor $\boldsymbol{\check{\sigma}}$ is applied over the outer boundary $\partial \Omega_{0}$ of a uniform elastic sphere.

Invoking Betti reciprocity to relate (2) to the auxiliary problem gives

\begin{equation}
    \int_{\partial \Omega_0}(\mathbf{E} \mathbf{x}) \cdot(\boldsymbol{\check{\sigma}} \mathbf{n})\,  \text{d} s-\int_{\partial \Omega_{\mathrm{inc}}}(\epsb^{I} \mathbf{x}) \cdot(\boldsymbol{\check{\sigma}} \mathbf{n})\,  \text{d} s = 
    \int_{\partial \Omega_0}(\mathbb{S}: \boldsymbol{\check{\sigma}}) \mathbf{x} \cdot \mathbf{0}\, \text{d} s-\int_{\partial \Omega_{\mathrm{inc}}}\left(\mathbb{S} :\boldsymbol{\check{\sigma}}\right) \mathbf{x} \cdot(\boldsymbol{\tilde{\sigma}} \boldsymbol{n})\, \text{d} s,
\end{equation}
where $\partial \Omega_{\mathrm{inc}}$ denotes the surface of the inclusion. Here, on $\partial\Omega_{\mathrm{inc}}$, $\mathbf n$ is taken to point
outward from the inclusion into the matrix. The first term on the left-hand side uses the ellipsoidal deformation assumption, and the first term on the right-hand side is zero because the correction
problem is traction-free on the outer boundary. Using the divergence theorem, this is equivalent to 

\begin{equation}
    \int_{\Omega_0}
    \operatorname{div}
    \left(
    \boldsymbol{\check{\sigma}}\mathbf E\mathbf x
    \right)\,\text{d}v
    -
    \int_{\Omega_{\mathrm{inc}}}
    \operatorname{div}
    \left(
    \boldsymbol{\check{\sigma}}\epsb^{I}\mathbf x
    \right)\,\text{d}v
    =
    -
    \int_{\Omega_{\mathrm{inc}}}
    \operatorname{div}
    \left[
    \tilde{\boldsymbol{\sigma}}
    (\mathbb S:\boldsymbol{\check{\sigma}})\mathbf x
    \right]\,\text{d}v .
\end{equation}
These integrals can be directly evaluated.  Upon simplifying, we obtain

\begin{equation}
    \left(
    R^3\mathbf E
    -
    ab^2\epsb^{I}
    +
    ab^2\mathbb S:\tilde{\boldsymbol{\sigma}}
    \right):\boldsymbol{\check{\sigma}}=0 
\end{equation}

for all $\boldsymbol{\check{\sigma}}$, meaning that

\begin{equation}
    \mathbf{u}^{\text{corr}}\left(\mathbf{x} \in \partial \Omega_0\right)=\frac{ab^2}{R^3}\left(\epsb^{I}-\mathbb{S}:\boldsymbol{\tilde{\sigma}}\right) \mathbf{x}.
    \label{eq:disp_field1}
\end{equation}

The prefactor on the right-hand side of Eq.~\eqref{eq:disp_field1} corresponds to the volume ratio between the inclusion and the sphere, and the term in parentheses is the difference between the strain inside the inclusion and the elastic strain associated with the compensating stress $\tilde{\boldsymbol{\sigma}}$. Using the expression for the inclusion strain $\epsb^{I}$ from Eqs.~\eqref{eq:eshelby2} and \eqref{eq:transformation_stress}, Eq.~\eqref{eq:disp_field1} further simplifies to

\begin{equation}
    \mathbf{u}^{\text{corr}}\left(\mathbf{x} \in \partial \Omega_0\right)=\frac{ab^2}{R^3} \epsb^{T}\mathbf{x}.
    \label{eq:disp_field2}
\end{equation}

Recall that $\epsb^{T}$ is uniquely determined by the known average stress $\tilde{\boldsymbol{\sigma}}$ through the Eshelby tensor, and therefore Eq.~\eqref{eq:disp_field2} gives a closed, first-order approximation for the inclusion-induced correction to the surface displacement field. It can account for an arbitrary number of contact forces, and could be extended to force distributions, as well as to non-ellipsoidal inclusions if an appropriate Eshelby-like tensor is available. It is also consistent in the limit where the inclusion volume is zero, or if there are no applied forces. For a fixed $\boldsymbol{\varepsilon}^{T}$, the normalized surface correction, $\mathbf{u}^{\mathrm{corr}}/R$, scales with the inclusion-to-sphere volume ratio $ab^2/R^3$.

A higher-order approximation of the surface displacement profile for problem (2) can be obtained by considering a surface displacement profile that includes variations beyond the linear form in Eq. \ref{eq:disp_field2}. For this higher-order correction, we specialize to an isotropic matrix with Young's modulus $E$ and Poisson ratio $\nu$. If the driving stress $\tilde{\sigb}$ and the inclusion geometry share principal directions, the natural next order correction, which respects the symmetry, is to consider the third-order form as shown below
\begin{equation}
    \mathbf{u}^{\text{corr}}\left(\mathbf{x} \in \partial \Omega_0\right)=\frac{a b^2}{R^3} \epsb^T \mathbf{x}+C\left[\frac{1}{R^2}\left(\mathbf{x} \cdot\left(\epsb^T\right)^{\prime} \mathbf{x}\right) \mathbf{x}-\frac{2}{5}\left(\epsb^T\right)^{\prime} \mathbf{x}\right]\label{highu}
\end{equation}
where $(\cdot)'$ denotes the deviator and $C$ is an as-yet undetermined coefficient.  This new expression also takes inspiration from the exact far-field solution to the original Eshelby problem\cite{Bower_2009}, which includes a cubic term in $\mathbf{x}$. The second term in the correction (linear in $\mathbf{x}$) ensures that the correction term does not affect the outcome of the previous linear analysis even if it were presumed a priori. To determine $C$, we may use reciprocity again, but this time using an auxiliary elastic solution whose boundary traction is sensitive to the higher-order correction above.

We may construct such an auxiliary solution using the Papkovich-Neuber formulation which instructs how to create stress and displacement solutions from a harmonic function.  Since we need a solution within a sphere and it must contain higher-order surface variations, we let our Laplace solution $w$ come from a third-order spherical harmonic to obtain $w(x_1,x_2,x_3)=5 x_3^3 - 3 r^2 x_3$ for $r=\sqrt{x_1^2+x_2^2+x_3^2}$.  We then construct the corresponding stress and displacement fields as 
\[\check{\sigma}_{ij}=\frac{2\nu}{2(1-\nu)}\frac{\partial\psi_k}{\partial x_k}\delta_{ij}+\frac{1-2\nu}{2(1-\nu)}\left(\frac{\partial\psi_i}{\partial x_j}+\frac{\partial\psi_j}{\partial x_i}\right)-\frac{x_k}{2(1-\nu)}\frac{\partial^2\psi_k}{\partial x_i \partial x_j}\, ,\]
\[\check{u}_i=\frac{2(1+\nu)}{E}\left(\psi_i-\frac{1}{4(1-\nu)}\frac{\partial}{\partial x_i}(x_k\psi_k)\right)\]
for vector potential $\boldsymbol{\psi}$ given by $\psi_1=\psi_2=0$ and $\psi_3=w$. Applying Betti reciprocity to relate problem (2) under the assumption of Eq. \eqref{highu} to the new auxiliary solution, $\check{\sigb}$ and  $\check{\mathbf{u}}$, we may now solve for $C$ to obtain
\[C=\frac{35 a b^2}{2(7+5\nu)R^3}.\]
With this value for $C$, the higher-order approximation, Eq. \eqref{highu}, is completely determined.

\subsection{Supplementary Note 4: Influence of inclusion geometry on architected grains}
\hfill\\
To systematically evaluate the role of inclusion geometry on grain-level mechanical response, we independently varied the orientation angle ($\beta$), aspect ratio ($a/b$), and relative density ($\bar{\rho}$) while maintaining constant overall grain dimensions (Fig. 2 and Extended Data Fig. 3). When the hollow elliptical void is aligned with the loading direction ($\beta = 0^\circ$), grains exhibit a softer initial response due to axial compliance of the elongated inclusion, promoting bending-dominated deformation and axial collapse. In contrast, transverse alignment ($\beta = 90^\circ$) results in higher stiffness at larger displacements as load transfer occurs across the minor axis, enhancing load-bearing capacity. The intermediate configuration ($\beta = 45^\circ$) produces a transitional behavior. Varying the aspect ratio ($a/b$ = 1.0, 1.5, 2.0) at constant mass further demonstrates that geometry governs the balance between compliance and nonlinear stiffening. Increasing $a/b$ reduces initial stiffness due to enhanced compliance along the major axis but leads to greater strain hardening at larger displacements as stress concentrates near the high-curvature ellipse tips. High-aspect-ratio inclusions exhibit more pronounced nonlinear behavior compared to circular voids ($a/b = 1.0$), arising from anisotropic stress redistribution and localized bending that promotes controlled strain localization and progressive load bearing rather than abrupt failure. Finally, decreasing relative density ($\bar{\rho}$ = 0.93, 0.83, 0.78) predictably reduces global stiffness and lowers force at a given displacement (Fig. 2). 

Together, these results define a three-parameter geometric design space in which orientation controls directional stiffness and deformation pathways, aspect ratio governs nonlinear compliance and strain localization, and relative density scales load-bearing capacity and energy absorption. Crucially, orientation and aspect ratio enable stiffness tuning at constant mass, effectively decoupling anisotropy from density. 

\subsection{Supplementary Note 5: Influence of inclusion geometry and confinement on architected granular packings}
\hfill\\
Quasi-static mechanical behavior of these architected granular packings was assessed via \emph{in situ} indentation using a blunt conical tip (Fig. 3 and Extended Data Fig. 4). It was observed that decreasing grain relative density consistently reduced the intrinsic stiffness of individual grains. However, unliked isolated grains, this stiffness reduction was partially offset in the packed configuration. Softer grains underwent greater lateral expansion under compression, thereby engaging more strongly with their neighbors and increasing inter-grain contact forces. Consequently, variations in inclusion aspect ratio and orientation, while significant in isolated grain experiments, had negligible influence under confined loading, where load transfer was dominated by inter-grain contacts, friction, and lateral constraint. 
Post-mortem X-ray computed tomography (XCT) revealed plastic deformation of the individual grains and local grain rotations, indicating that both deformation of the grain walls and inter-granular friction contribute to energy dissipation (Supplementary Fig. 2). These observations highlight a clear confinement-dependent dissipation mechanism: In 3D wells, tight packing constraints suppress global rearrangement, making plastic deformation of the individual grains the dominant energy loss pathway, with minor contributions from local grain rotation. In
2D wells, reduced confinement enables global rearrangement and stuctural reconfiguration prior to grain compression, resulting in a more balanced contribution from plastic deformation of the individual grains and frictional sliding (Supplementary Fig. 2). 

\subsection{Supplementary Note 6: Design of granular-inspired midsole and cushion}
\hfill\\
granular-inspired metamaterials at the macroscale offer a broad range of applications and have the potential to replace conventional lattice materials across diverse disciplines. For example, 2D granular metamaterials (Fig.~5a) can be tessellated with spatial variations in fill fraction, eccentricity, and orientation to construct footwear midsoles (Fig.~5c). Unlike conventional lattice- or foam-based midsoles—where geometric compatibility between adjacent building blocks must be carefully satisfied when introducing spatial heterogeneity—our granular-inspired metamaterials enable more effective programming of stiffness and dissipation by tuning the fill fraction and eccentricity within each grain, without compatibility constraints. This capability allows precise control over the spatial distributions of both mass density and stiffness within the metamaterial midsole. To realize this design, we first built up a negative mold comprising spatially distributed elliptical inclusions, (Extended Data Fig. 9a). The geometry of each elliptical inclusion is defined as a function of the representative volume element (RVE) indices $(i,j)$ (Extended Data Fig. 9a,b). The negative mold consists of 24 and 90 RVEs along the vertical and horizontal directions, respectively. The semi-axis lengths of the inclusions, $a_{ij}$ and $b_{ij}$, are given by Eqs.~\ref{eq:midsole_aij} and \ref{eq:midsole_bij}, where $R$ denotes the grain radius. After generating the graded negative mold, the granular metamaterial midsole is obtained by performing a Boolean subtraction of the mold from the midsole domain (Extended Data Fig. 9c).
\begin{equation}
a_{ij}=
\begin{cases}
0.43R, & i < 18,\\[8pt]
\left(1-\dfrac{i-18}{12}\right)0.43R
+\dfrac{i-18}{12}\,(0.30R), 
& 18 \le i \le 30,\\[12pt]
0.3R(1-\frac{i-30}{60})\,, & 30<i \le 90.
\end{cases}
\label{eq:midsole_aij}
\end{equation}

\begin{equation}
b_{ij}=
\begin{cases}
0.70R, & i < 18,\\[8pt]
\left(1-\dfrac{i-18}{12}\right)0.70R
+\dfrac{i-18}{12}\,(0.30R), 
& 18 \le i \le 30,\\[12pt]
0.3R(1-\frac{i-30}{60})\,, & 30<i \le 90.
\end{cases}
\label{eq:midsole_bij}
\end{equation}
The spatial heterogeneity can also be realized in a 3D tessellation to build a metamaterial cushion, where the ellipsoidal void of each grain can manifest its own size and orientation, as demonstrated in Fig.~\ref{fig:metamaterial_cushion}; this programmability enables a tunable stiffness distribution via altering the internal geometry of individual grains without modifying inter-grain connections.

\subsection{Supplementary Note 7: Comparison between granular-inspired metamaterials and lattice materials}
\hfill\\
In contrast to conventional lattice-based materials, which dissipate impact energy mainly through the viscosity and plasticity of their constituent materials, granular-inspired materials introduce an additional dissipation mechanism: inter-grain friction under large deformation. Extended Data Fig. 8 compares the force mitigation and energy dissipation responses of a representative granular-inspired material and several conventional lattice materials with identical mass. For periodic lattice materials, including triangular, square, and hexagonal lattices, the maximum impact force is positively correlated with the lattice coordination number. The corresponding post-impact rebound velocities of the projectile are 0.49 m/s, 0.43 m/s, and 0.57 m/s for the triangular, square, and hexagonal lattices, respectively. In contrast, the aperiodic Voronoi lattice results in a substantially lower impact force because of its structural heterogeneity, with a rebound velocity of 0.45 m/s. Unlike these lattice materials, which fail to reconcile force mitigation and energy dissipation within a single architecture, the granular-inspired material, shown in the left column of Extended Data Fig. 8, exhibits a maximum impact force of 1.30 kN and a rebound velocity of only 0.22 m/s. This demonstrates its ability to achieve both moderate force mitigation and substantial energy dissipation.

\subsection{Supplementary Note 8: Design of spatially programmable contact networks}
\hfill\\
Spatially localized contact activation within the granular metamaterial is achieved by encoding spatial variations in strut geometry at the grain level, thereby prescribing regions of preferential deformation and inter-grain engagement under compression (Extended Data Fig. 10). The underlying design principle exploits the sensitivity of slender structural elements to buckling. Specifically, the critical buckling load of a strut scales with its bending stiffness, which depends on both material modulus and geometric moment of inertia. By locally reducing strut thickness within selected regions of the grain architecture, the corresponding elements exhibit reduced buckling resistance and therefore undergo preferential instability under global loading. This controlled instability leads to localized collapse and increased inter-grain contact in targeted zones (i.e., compliant regions), effectively activating contact pathways in a spatially prescribed manner. In contrast, regions with thicker struts maintain higher bending stiffness and remain comparatively stable over the same loading range, thereby suppressing premature deformation and limiting contact formation outside the intended zones (i.e., stiff and interface regions). This stiffness contrast enables a deterministic partitioning of the structure into highly deformable and minimally deformable regions, independent of the global packing configuration. More complex contact topologies are achieved by simultaneously modulating strut thickness across the architecture. By selectively thinning struts in regions designated for activation while thickening neighboring regions, deformation can be redistributed and guided along predefined pathways. This approach suppresses unintended strain localization and prevents the formation of uncontrolled contact networks, ensuring that inter-grain engagement occurs predominantly along the encoded geometry. 

The resulting contact activation process can thus be understood as a geometry-programmed instability, in which local variations in structural stiffness dictate the sequence and spatial distribution of buckling events. Importantly, this mechanism decouples contact network formation from packing disorder and boundary conditions, enabling the design of reproducible and spatially controlled contact pathways across the material. These design principles are general and can be extended to different geometries, length scales, and material systems, provided that sufficient contrast in local structural stiffness is introduced to bias deformation. In practice, the magnitude of the stiffness contrast required for robust localization depends on the relative geometry of the struts and the applied loading conditions, but can be systematically tuned through parametric variation of strut thickness and connectivity.

\newpage
\subsection{Supplementary Figures}
\hfill\\
 \begin{figure}[H]                  
    \captionsetup{labelformat=supplementary,labelsep=vline}
    \centering
    \includegraphics[width=1.0\textwidth]{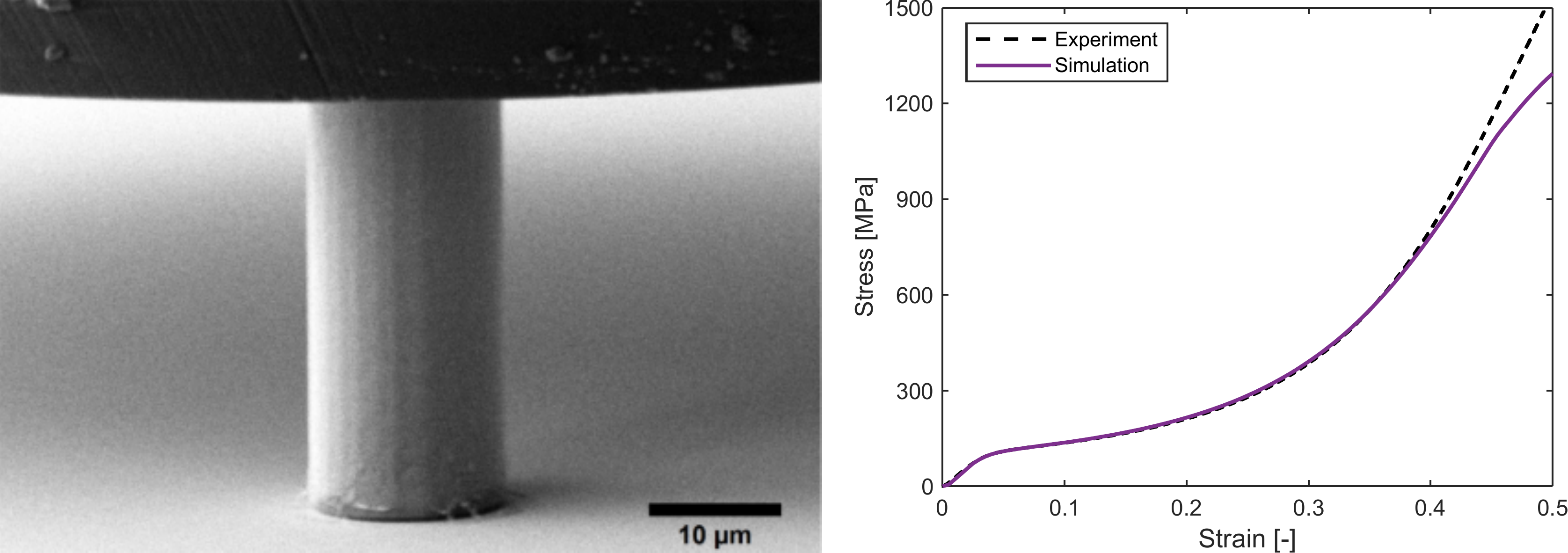}
    \caption{\textbf{Constituent material characterization.}
    Uniaxial compression response of the two-photon polymerized constituent material used in microscale grains. The extracted elastic–plastic properties were used as input parameters for nonlinear finite element simulations.}
    \label{fig:s1_constituent}
\end{figure}

\newpage
 \begin{figure}[H]
    \captionsetup{labelformat=supplementary,labelsep=vline}
    \centering
    \includegraphics[width=1.0\textwidth]{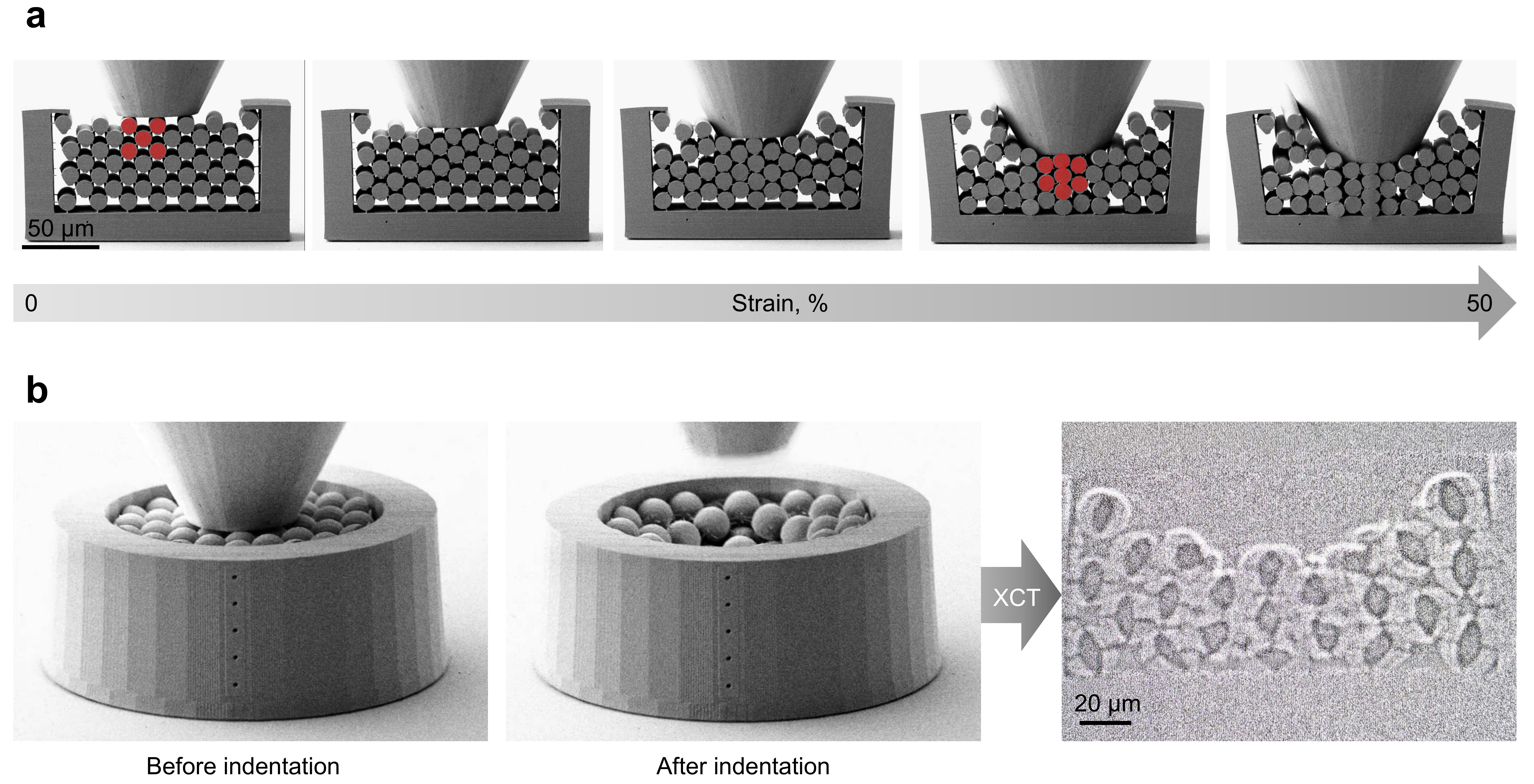}
    \caption{\textbf{Deformed morphologies of granular packings.} \textbf{a}, \emph{In situ} indentation of 2D microscale FCC granular packing, showing global grain rearrangement and sliding. \textbf{b}, Cross-section XCT reconstruction of a deformed 3D microscale granular packing with architected grains after indentation showing plastic deformation and local rotation of the grains.}
    \label{fig:s3_XCT_packing} 
\end{figure}

\newpage
 \begin{figure}[H]
    \captionsetup{labelformat=supplementary,labelsep=vline}
    \centering
    \includegraphics[width=0.96\textwidth]{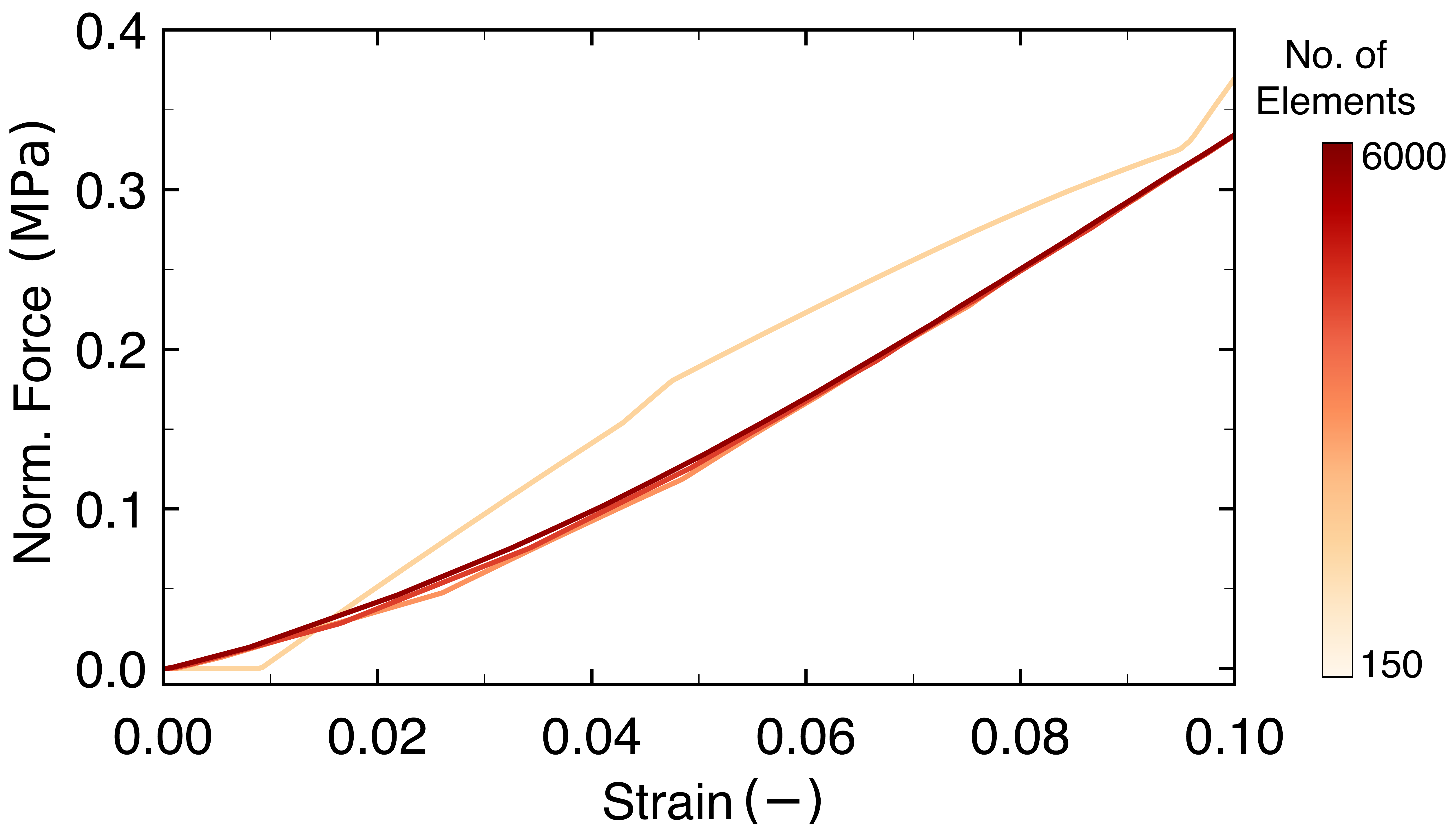}
    \caption{\textbf{Mesh convergence and contact sensitivity in FEM simulations.}}
    \label{fig:pillar} 
\end{figure}

\newpage
 \begin{figure}[H]
    \captionsetup{labelformat=supplementary,labelsep=vline}
    \centering
    \includegraphics[width=1.0\textwidth]{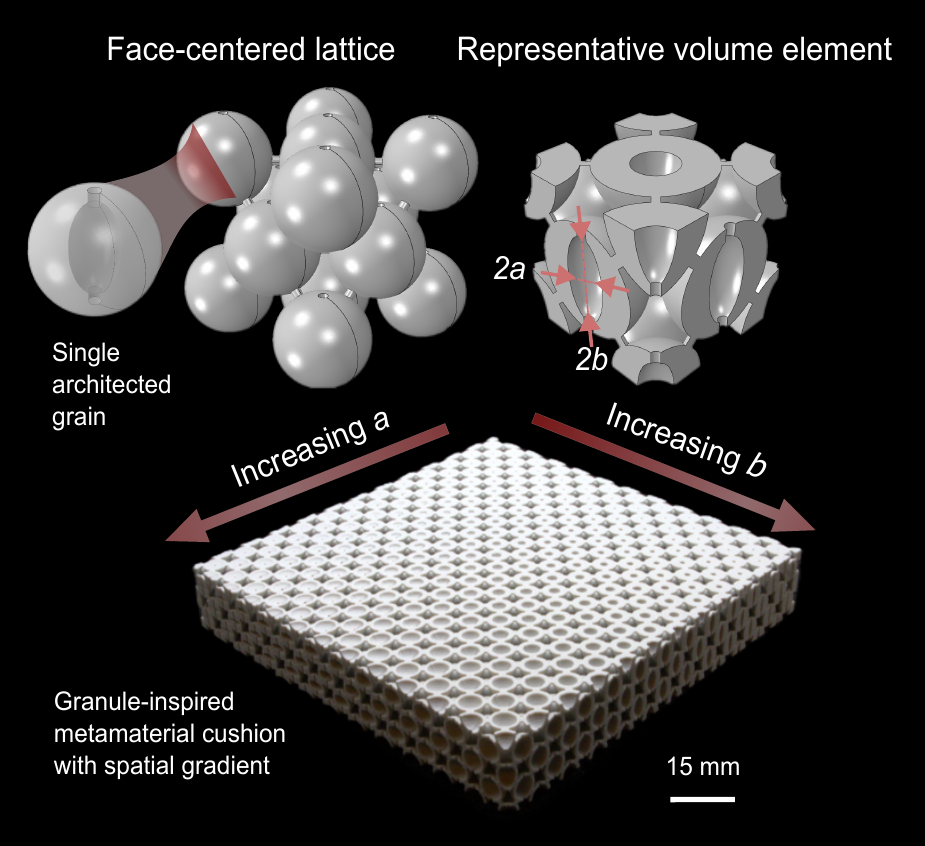}
    \caption{\textbf{Design of granular-inspired cushion with spatial gradient.} }
    \label{fig:metamaterial_cushion} 
\end{figure}

\newpage
\subsection{Supplementary video captions}
\hfill

\noindent\textbf{Supplementary video 1.} \\
\hspace{1em}\emph{In situ} SEM compression of dense versus architected grain. SEM video showing lateral expansion of an architected grain with a hollow elliptical inclusion during quasi-static compression, highlighting enhanced transverse deformation relative to a fully dense grain.

\noindent\textbf{Supplementary video 2.} \\
Quasi-static deformation response of 2D and 3D granular packings under indentation.

\noindent\textbf{Supplementary video 3.} \\
High-rate particle impact on granular metamaterial. High-speed imaging of silica microparticle impact during LIPIT experiments, showing reduced rebound velocity and increased impact duration in low-density architected packings.

\noindent\textbf{Supplementary video 4.} \\
Simulated particle impact on 3D granular packings with fully dense and architected grains visualizing grain-scale deformation, contact activation, and force redistribution during dynamic impact.

\noindent\textbf{Supplementary video 5.} \\
Macroscale impact response of granular-inspired lattices. Drop-tower impact experiment comparing rebound height and force–time response of architected granular lattices and conventional triangular lattices.

\nolinenumbers
\vspace{10pt}

\end{document}